\documentclass[]{aa}  

\usepackage{graphicx}
\usepackage{txfonts}
\usepackage{amsmath}
\usepackage{xcolor}
\begin{document}

   \title{Residual eccentricity of  an Earth-like planet  \\
   orbiting a red giant Sun}


   \author{A.~F.~Lanza
          \inst{1}
          \and
          Y.~Lebreton\inst{2,3}
          \and
          C.~Sallard\inst{4}
          }

   \institute{INAF-Osservatorio Astrofisico di Catania, Via S.~Sofia,78 - 95123 Catania,Italy\\
              \email{antonino.lanza@inaf.it}
         \and
            LESIA, Observatoire de Paris, Universit\'e PSL, CNRS, Sorbonne Universit\'e, Universit\'e de Paris, 5 Place Jules Janssen, 92195, Meudon, France 
            \and
            Universit\'e de Rennes, CNRS, IPR (Institut de Physique de Rennes) -- UMR 6251, 35000, Rennes, France
          \and
          Ecole Universitaire de Physique et d'Ingénierie, Université Clermont-Auvergne, Site des C\'ezeaux, Clermont-Ferrand, France
             }

   \date{Received ...; accepted ...}

 
  \abstract
   {The late phases of the orbital evolution of an Earth-like planet around a Sun-like star are revisited considering the effect of the density fluctuations associated with convective motions inside the star. }
   {Such fluctuations produce a random perturbation of the stellar outer gravitational field that excites a small residual eccentricity in the orbit of the planet counteracting the effects of tides that tend to circularize the orbit. }
   {We compute the power spectrum of the outer gravitational field fluctuations of the star in the quadrupole approximation and study their effects on the orbit of the planet using a perturbative approach. The residual eccentricity is found to be a stochastic variable showing a Gaussian distribution.   }
   {Adopting a model of the stellar evolution of our Sun computed with MESA, we find that the Earth will be engulfed close to the tip of the red giant branch evolution phase. We find a maximum mean value of the residual eccentricity of $\sim 0.026$ immediately before the engulfment.  Considering an Earth-mass planet with an initial orbital semimajor axis sufficiently large to escape engulfment, we find that the mean value of the residual eccentricity is greater than 0.01 for an initial separation up to $\sim 1.4$~au.}
   {The engulfment of the Earth by the red giant Sun is found to be a stochastic process instead of being deterministic as assumed in previous studies.  If an Earth-like planet escapes engulfment, its orbit around its remnant white dwarf star will be moderately eccentric. Such a residual eccentricity on the order of a few hundredths can play a relevant role in sustaining the pollution of the white dwarf atmosphere by asteroids and comets as observed in several objects. }


   \maketitle
%

\section{Introduction}
Red giant stars with an initial mass similar to that of the Sun reach a luminosity of thousands of solar luminosities and a radius of a few hundreds solar radii close to the tip of the red giant branch \citep[RGB, e.g.][]{KWW12}.  Convective motions inside the extended envelopes of such stars transport virtually all those large luminosities from the energy generation layers up to their photospheres. Density fluctuations are associated with convection because rising hotter columns of plasma are less dense than the average, while cool descending columns are denser. Such density fluctuations with a relative amplitude up to $\approx 10^{-5}$ produce a slightly fluctuating outer gravitational potential of the red giant that can affect the orbit of a close-by companion. This effect was proposed by \citet{Phinney92} and \citet{PhinneyKulkarni94} to account for the non-zero eccentricity of binary millisecond pulsars where very small eccentricities ($e \la 10^{-5}$) can be measured thanks to the exquisite precision in the measurement of the orbital Doppler shift  made possible by the periodic pulses coming from the rapidly spinning neutron star. 

According to these authors, such low-eccentricity millisecond pulsars are the end products of low-mass X-ray binaries the donor star of which evolves off the main sequence becoming a red giant, thus starting to transfer mass and angular momentum to a very old neutron star companion. This produces the X-ray emission and spins up the neutron star to a rotation period in the millisecond range. Once the mass transfer ceases, we observe the system as a millisecond binary pulsar at radio wavelengths \citep[see, e.g.,][for details and discussion of the different formation channels for those systems]{TaurisSavonije99,TaurisvandenHeuvel06,Chenetal21}.

The evolution phase of such systems relevant for our consideration occurs when the red giant approaches the end of its nuclear fuel reservoir and starts to contract detaching from its Roche lobe, thus ending the mass transfer to the neutron star. In such a phase, the orbit of the pulsar evolves under the action of tides in the contracting red giant and the density fluctuations in its convective envelope that produce a slightly fluctuating outer gravitational potential. This induces a small eccentricity in the orbit of the binary that tides are not capable of completely erasing because the radius of the secondary is drastically reduced on a timescale shorter than the damping  timescale of the residual eccentricity. Such a residual eccentricity, therefore, remains as a relic of the final phase of the evolution of the detached secondary before it becomes a white dwarf. It is a stochastic variable because it is produced by the random density fluctuations inside the convective envelope of the contracting red giant \citep[cf.][]{Phinney92}. 

The computation of the residual eccentricity in the case of millisecond pulsar binaries was revisited by \citet{LanzaRodono01} who applied the hydromagnetic virial theorem of  \citet{Chandrasekhar61} and methods of the stochastic differential equation theory to compute its statistical distribution. Moreover, they suggested the relevance of the residual eccentricity for planets orbiting red giant stars, but did not investigate the topic \citep[see a parallel treatment in Sect.~5 of][]{KissinThompson15}.  

{  Because the Sun will become a red giant, such an investigation is relevant to predict the late phases of the evolution of our planetary system.  
The evolution of the orbits of the inner solar system planets, including the Earth, during the remaining lifetime of the  Sun on the main sequence, cannot be predicted in a deterministic way because of the chaotic character of their long-term dynamics \citep[e.g.,][]{Laskar96,MogaveroLaskar21}. Numerical integrations of thousands of solar system  realizations starting from its present dynamical state within the range of its uncertainty or simplified analytical models show that the planet Mercury has a probability on the order of $0.5-1$\% of colliding with Venus or with the Sun within the next 5~Gyr \citep{LaskarGastineau09,Batyginetal15}. While most of those Mercury's highly 
 eccentric orbits do not destabilize the inner solar system, a small fraction of them could result in the excitation of large changes in the orbit eccentricities of the Earth, Mars or Venus leading to close encounters or collisions between these planets. The probability of such a catastrophic outcome for our planet before the Sun  leaves the main sequence is difficult to estimate, but it is likely to be below 0.1\% \citep[e.g.,][]{LaskarGastineau09,Zeebe15}. Therefore, in our subsequent considerations, we shall assume that the eccentricity of the Earth orbit is not significantly modified by the perturbations of the other planets during the entire lifetime of the solar system. In other words,  the evolution of the Sun appears us to be the most relevant cause of changes in the orbit of the Earth and will most likely determine its final fate. 

\citet{Sackmannetal93} explored the evolution of our star in detail and considered the orbital expansion produced by the solar mass loss because of the conservation of the orbital angular momentum (cf. Sect.~\ref{mass_loss_and_tides}), but did not include the effects of tides that counteract it by producing a decay of the semimajor axis and damp any initial eccentricity of the orbits.  
\citet{RybickiDenis01} re-examined the solar system's fate by discussing  several other models for the solar evolution and including tidal  effects. They showed that the drag exerted by the solar wind, the mass accretion from the wind, and the evaporation of our planet under the solar irradiation have negligible effects on the evolution of the Earth's orbit, even during the RGB phase and up to the tip of asymptotic giant branch (AGB) phase. Therefore, the most relevant effect in determining whether the Earth will be finally engulfed by the Sun or not appears to be the solar mass loss rate because it rules both the largest radii reached on the RGB and AGB phases and the orbit expansion \citep[see also][]{Guoetal16}. 
\citet{SchroederSmith08} adopted a specific model to predict the solar mass loss and found that the Sun  will reach its maximum radius at the end of the RGB phase. Including the effects of tides, they found that the Earth will not be able to escape engulfment at the tip of the RGB phase despite the expansion of its orbit. }

The consideration of the residual eccentricity of the Earth's orbit, produced by the density fluctuations in the extended envelope of the red giant Sun, will add an important conceptual difference to the final phases of our planet orbital evolution. While, in principle, the time of the Earth engulfment can be predicted by models such as those of \citet{RybickiDenis01} or \citet{SchroederSmith08}, provided that one has both accurate stellar evolution model and tidal theory, this is no longer possible with the introduction of the residual eccentricity because it is a stochastic variable for which we can only compute a statistical probability distribution. The difference from a practical point of view is negligible because the final fate of the Earth will not change, but the residual eccentricity may play a role in the final phases of the dynamical evolution of the solar system because it can induce secular changes in the eccentricities of the orbits of small bodies such as asteroids or comets. This is not possible when the orbit of the Earth is circularized during the late phase of its evolution as predicted on the basis of the tidal theory alone. Putting an asteroid on a highly eccentric orbit may produce its capture by the white dwarf Sun leading to its fall on the degenerate star, an event that will pollute its atmosphere with metals as observed {  in at least  $\sim 25$\% of white dwarfs \citep[cf.,][and references therein, especially the list in their Fig.~6]{FrewenHansen14,Veras21}}. 

In the present work, we determine the statistical distribution of the residual eccentricity of the orbit of an Earth-like planet along the evolution of a Sun-like star up to its final phases. Moreover, if the planet is transiting across the disc of a red giant, the  fluctuation of its outer gravitational potential affects  the timing of the transits in a potentially measurable way. We revisit the methods to compute the residual eccentricity and the transit time variations developing an approach simpler than that of  \citet{LanzaRodono01} that allows a straightforward derivation of all the relevant equations and include the effects of the stellar mass loss and tides during the evolution of the star in its red giant phase. Our model to compute the statistical distribution of the residual eccentricity and the transit time variations is presented in Sect.~\ref{model}, while an internal structure and evolution model of a Sun-like star is introduced in Sect.~\ref{stellar_str_evol_model}, and an illustrative application to an Earth-Sun-like system is made in Sect.~\ref{applications}. We discuss our results and conclude in Sect.~\ref{conclusions}. 

\section{Model}
\label{model}
In this section, we introduce the model to compute the residual eccentricity and the transit time fluctuations, while its application to an Earth-like planet in orbit around a Sun-like star will be presented in Sect.~\ref{applications}. 
The effects of the stellar mass loss and of the tides on the planet orbit are discussed in Sect.~\ref{masslossandtides}, while the computation of the outer gravitational potential of the star is presented in Sect.~\ref{grav_poten}. The variations in the orbital elements of the planet under the perturbations produced by such a fluctuating potential are introduced in Sect.~\ref{perturbations}, separately for the eccentricity (Sect.~\ref{eccentr_var}) and the mean longitude at the epoch from which the transit time variations can be derived (Sect.~\ref{mean_long_var}). The power spectrum of the gravitational potential fluctuations, required to compute the fluctuations in the orbital elements,  is computed in Sect.~\ref{power_sp}. 

{  The evolution of the orbit of a planet around a RGB star has been investigated in order to account for the observed lack of giant planets with a semimajor axis shorter than $\sim 0.55$~au around giant stars \citep[cf. Sect.~4 of][]{Villaveretal14}. The maximal initial orbital semimajor axis that ensures the survival of a planet of given mass has been determined under the effects of tides and stellar mass loss considering stars of different masses and adopting different prescriptions for their mass loss rates and several evolution codes. Initially, most of the investigations have been limited to solar-like or intermediate-mass stars evolving from the main-sequence up to the RGB  \citep[e.g.][]{VillaverLivio09,Kunimotoetal11,Villaveretal14}. Subsequent works have investigated the fate of planets around AGB stars, including the thermal pulse phase \citep{MustillVillaver12} and considering also companions up to the stellar mass range \citep[e.g.][]{NordhausSpiegel13} in an attempt to predict  the formation of planetary or stellar binary systems with a white dwarf central star.  Similar models have been computed to predict the fraction of planetary nebulae with central binary systems \citep[cf.][]{Madappattetal16}. Most of those studies have focused on systems consisting only of a single planet \citep[see, however, the discussion in \S~5.2 of][]{MustillVillaver12}, while an investigation of the role of the mutual gravitational perturbations in a system consisting of an inner Neptune-mass and an outer Jupiter-mass planets around a red giant star has been pursued by \citet{Roncoetal20}. 

The most important effects affecting the orbits of late planets have been found to be the stellar mass loss and the tides raised by the planets inside their giant host stars \citep[see][for an analytic model of their joint effects]{AdamsBloch13}. On the other hand, the impact of giant planets on the evolution of their host stars, in particular on their angular momentum, mass loss, and hydromagnetic dynamo action, has also been investigated \citep[e.g.][]{SokerHarpaz00,Carlbergetal09,Priviteraetal16a,Priviteraetal16b,Priviteraetal16c}.

In the present investigation, we shall introduce a model to compute the residual eccentricity and the transit time fluctuations and show a simple application to the case of a system consisting of a single Earth-mass planet orbiting a Sun-like star because a general  investigation of the fates of planets and planetary systems around RGB and AGB stars is well beyond the scope of the present work. Therefore, we refer the interested reader to the above papers for information on such a wider scenario. }


\subsection{Stellar mass loss and tides}
\label{mass_loss_and_tides}
Let us consider a system consisting of a planet of mass $m_{\rm p}$ orbiting a Sun-like star of mass $m_{\rm s}$; the semimajor axis of the orbit is $a$, while its eccentricity is $e$. The orbital angular momentum is given by:
\begin{equation}
J = m \sqrt{G m_{\rm T} a (1 - e^2)} = m n a^{2} \sqrt{1 -e^{2}},
\label{orbital_ang_mom}
\end{equation}
where $m\equiv m_{\rm p} m_{\rm s}/m_{\rm T}$ is the reduced mass of the system, $m_{\rm T} \equiv m_{\rm s} + m_{\rm p}$ its total mass, and we made use of Kepler third law:
\begin{equation}
 n^{2} a^{3} = G m_{\rm T},
\end{equation}
where $n= 2\pi/P_{\rm orb}$ is the orbital mean motion with $P_{\rm orb}$ being the orbital period. Given than $m_{\rm p} \ll m_{\rm s}$ and the orbit is nearly circular ($e \ll 1$), the conservation of the orbital angular momentum allows us to find the variation in the orbit semimajor axis produced by a variation in the mass of the star as
\begin{equation}
\frac{1}{a} \frac{da}{dt} \simeq - \frac{1}{m_{\rm s}} \frac{d m_{\rm s}}{dt}, 
\label{dadt0}
\end{equation}
that can be used to compute the dynamical effect of the star mass loss \citep[cf.][Sect.~2]{SchroederSmith08}. 

{Equation~\eqref{dadt0} is obtained by considering an isotropic stellar mass loss \citep{Verasetal11}. Rigorously speaking, any  mass loss perturbs the orbit of the star-planet system producing variations in its orbital elements that can no longer be regarded as constant as in the case of the standard two-body problem. Osculating orbital elements can be introduced to describe the instantaneous orbit under the effects of the mass loss perturbation \citep[e.g.][]{DosopoulouKalogera16a}. From a dynamical point of view, the mass loss regime can be characterized through the parameter 
\begin{equation}
\Psi_{\rm ml} \equiv \frac{1}{n m_{\rm T}}\frac{dm_{\rm T}}{dt}  \simeq \frac{1}{n m_{\rm s}}\frac{dm_{\rm s}}{dt},  
\label{Psi_ml_eq}
\end{equation}
introduced by, e.g., \citet{Verasetal11}. When $\Psi_{\rm ms} \ll 0.1 $, the mass loss timescale is much longer than the orbital period and the system is in the so-called "adiabatic mass loss regime" that allows us to average the variations in the osculating elements along the orbit and obtain variation equations that are independent of the orbit true anomaly. It is in that regime that Eq.~\eqref{dadt0}, rigorously speaking, is valid. It is interesting to note that in the case of an anisotropic stellar mass loss in the adiabatic regime, the semimajor axis remains secularly constant because the effects of the anisotropic mass flux average to zero \citep{Verasetal13,DosopoulouKalogera16b}. Therefore, only the isotropic mass loss is to be included into Eq.~\eqref{dadt0}. }

{  The variation in the eccentricity due to an isotropic stellar mass loss, averaged along the orbit, becomes zero when computed in the  adiabatic regime. Nevertheless, the eccentricity as an osculating element oscillates along the orbit with an amplitude of $\Psi_{\rm ml} (1-e^{2})$, where $e$ is the eccentricity in the absence of mass loss \citep[cf.][Sect.~2.4.1]{Verasetal11}. Even assuming a mass loss of $0.1$~M$_{\odot}$ in 1~Myr by a solar-mass star and a planet with an orbital period of 2~years, the parameter $\Psi_{\rm ml} \sim 1.3 \times 10^{-6}$ and the oscillation of the eccentricity along the orbit is negligible in comparison with the residual eccentricity induced by convective fluctuations in the stellar envelope that reaches values up to $\sim 10^{-2}$ in the late stages of stellar evolution (cf. Sect.~\ref{applications}). Therefore, we can safely assume for our purposes that the eccentricity is not significantly excited by an isotropic mass loss from the star.   }

{  The second contribution to be considered in modeling the evolution of the orbit is the tidal torque that affects both the semimajor axis $a$ and the eccentricity $e$. } The tidal torque  scales as $(R_{\rm s}/a)^{6}$, where $R_{\rm s}$ is the radius of the star, therefore, tides will be relevant only after the star has started to ascend the RGB because only in that phase its radius will become sufficiently large to produce a significant tidal torque in the case of an Earth-mass planet on an Earth-like orbit ($a \sim 1$~au). Following \citet{Zahn89}, we write the tidal torque $\Gamma$ acting on the planetary orbit by the equilibrium tide as \citep[cf.][]{SchroederSmith08}
\begin{equation}
\Gamma = 6 \frac{\lambda_{2}}{t_{\rm f}} \left(\frac{m_{\rm p}}{m_{\rm s}} \right)^{2} m_{\rm s} R_{\rm s}^{2} \left( \frac{R_{\rm s}}{a} \right)^{6} \left( \Omega_{\rm s} - n \right),
\label{tidal_torque_zahn}
\end{equation}
where $\lambda_{2} \simeq 0.019\,  \alpha^{4/3}$ and $t_{\rm f} = (m_{\rm s} R_{\rm s}^{2}/L_{\rm s})^{1/3}$ are a nondimensional parameter and the convective friction timescale appearing in the equilibrium tide theory of \citet{Zahn77,Zahn89}, respectively, while $\Omega_{\rm s}$ is the angular velocity of rotation of the star, $L_{\rm s}$ its luminosity, and $\alpha$ the ratio of the mixing length to the pressure scale height. 

A Sun-like star on the RGB can be assumed to be practically non-rotating because of the loss of angular momentum during the evolution on the main-sequence and the sub-giant phases and the large increase of its moment of inertia due to the  radius expansion. In other words, we can assume $\Omega_{\rm s} \sim 0$ for our purposes, thus $\Gamma < 0$ which produces a decrease of the orbit semimajor axis. Considering a planet with an orbital period longer than $\sim 150-200$ days, the frequency of the equilibrium tide is smaller than the turnover frequency of convective eddies inside the star which justifies the use of an unreduced turbulent viscosity in the convective envelope when estimating the tidal dissipation \citep{Zahn89}. Moreover, our simplified expression of $\lambda_{2}$ in comparison with that in Equation~(15) of \citet{Zahn89} is justified in the case of an orbital period around one year because $t_{\rm f} \approx 1$ year for the Sun close to the tip of the RGB. The effects of dynamical tides can be neglected for a solar mass star when the orbital separation is larger than $0.2-0.3$~au as in our case \citep[see][for a discussion of the role of dynamical tides]{Raoetal18,Sunetal18}. 

Adding the effect of the tidal torque to that of the stellar mass loss, equation~(\ref{dadt0}) becomes
\begin{equation}
\frac{1}{a} \frac{da}{dt} \simeq -\frac{1}{m_{\rm s}} \frac{d m_{\rm s}}{dt} + \frac{2}{J} \, \Gamma,
\label{dadt_complete}
\end{equation}
that we shall use to compute the evolution of the semimajor axis of the planetary  orbit in our application in Sect.~\ref{applications}. 

Another effect of the equilibrium tide inside the red giant star is the damping of the orbital eccentricity of the planet. We shall compute the circularization time $\tau_{\rm e}$ according to \citet{VerbuntPhinney95} who calibrated the tidal theory of Zahn by means of a sample of binaries containing a giant component. Specifically, we shall adopt
\begin{equation}
\frac{1}{\tau_{\rm e}} \equiv -\frac{1}{e} \left( \frac{de}{dt} \right) = 
\beta \left( \frac{L_{\rm s}}{m_{\rm env} R_{\rm s}^{2}} \right)^{1/3} \left( \frac{m_{\rm env}}{m_{\rm s}}\right) \left( \frac{m_{\rm p}}{m_{\rm s}} \right) \left( \frac{m_{\rm T}}{m_{\rm s}} \right) \left( \frac{R_{\rm s}}{a} \right)^{8}, \label{tau_ecc}
\end{equation}
where $\beta$ is a non-dimensional empirical factor of order unity and $m_{\rm env}$  the mass of the convective envelope of the giant star where the kinetic energy of the tidal flow is dissipated. By fitting their binary sample,   \citet{VerbuntPhinney95} found $0.5 \la \beta \la 2$, that is, Eq.~\eqref{tau_ecc} provides an estimate of $\tau_{\rm e}$ within approximately a factor of two when we adopt $\beta =1$.  

{  Equation~\eqref{tau_ecc} does not include the variation in the eccentricity produced by an anisotropic stellar mass loss, an effect that cannot be excluded in the case of solar-like stars in the late stages of their evolution \citep[e.g.][]{Soker98}. Its impact on the eccentricity depends on the specific geometry and the time dependence of the mass loss. Several idealized cases have been explored by, e.g., \citet{Verasetal13} or \citet{DosopoulouKalogera16b}. The most relevant occurs when there is a longitudinal anisotropy in the mass loss because of a difference in the mass-loss rate between the stellar hemispheres of up to 1\% with an average mass loss rate of $\sim 3 \times 10^{-7}$~M$_{\odot}$~yr$^{-1}$. In this case, the anisotropy can excite an eccentricity on the order of 0.01 \citep[e.g.][Sect.~3.1.3]{Verasetal13}. Such an eccentricity adds to the residual eccentricity that we are considering in the present work and should be included as an additional effect when treating stars with an anisotropic stellar mass loss. 

In addition to the effects of stellar mass loss and tides, the orbit of the planet can be affected by the frictional and gravitational drag forces. Frictional drag occurs because the planet is an obstacle that moves inside the stellar wind, while the gravitational drag arises because of the gravitational field of the planet acting on the wind itself. Their effect is small in comparison to those of the tides and stellar mass loss \citep[cf.][]{DuncanLissauer98,VillaverLivio09,MustillVillaver12,Villaveretal14,Raoetal18,Yarzaetal22}, therefore we shall neglect them in our simplified model and  assume that the mass of the planet is constant throughout its evolution. This amounts to neglect planetary evaporation and mass accretion by the stellar irradiation and the stellar wind, respectively.}

\label{masslossandtides}

\subsection[Gravitation potential]{Gravitational potential of a non-spherically symmetric star}
\label{grav_poten}

The gravitational potential $\Phi$ of a star at an external point $P$ up to the quadrupole order can be expressed as:
\begin{equation}
\Phi = -\frac{Gm_{\rm s}}{r} - \frac{3G}{2r^{3}} \sum_{i, k } \frac{Q_{ik} x_{i} x_{k}}{r^{2}},
\label{ogp}
\end{equation}
where $G$ is the gravitation constant, $m_{\rm s}$ the mass of the star, $r > R_{\rm s}$ the distance of $P$ from the barycenter $O$ of the star, $Q_{ik}$ the quadrupole moment tensor of the star, and $x_{i}$ the coordinates of $P$ in an orthogonal Cartesian reference frame of origin $O$, while the indexes $i,  k = 1,2,3$ specify the Cartesian coordinates. 
The quadrupole moment tensor can be expressed in terms of the inertia tensor of the mass distribution of the star as
\begin{equation}
Q_{ik} = I_{ik} - \frac{1}{3} \delta_{ik} {\rm Tr}\, { I},
\label{qdef}
\end{equation}
where $\delta_{ik}=1\mbox{ for $i=k$ and $\delta_{ik}=0$ for $i\not= k$}$ is the Kronecker $\delta$ tensor and ${\rm Tr}\, {I} = I_{xx} + I_{yy} + I_{zz}$ is the trace of the inertia tensor $ I$, that is, the sum of its diagonal components. The components of the inertia tensor are given by   
\begin{equation}
I_{ik} = \int_{V} \rho({\vec  r}) x_{i} x_{k} \, dV,
\label{inerdef}
\end{equation}
where $\rho$ is the density, ${\vec  r} \equiv (x_{1}, x_{2}, x_{3})$ the position vector, and $V$ the volume of the star over which the integration is extended. 

The density inside a stellar convective envelope can be written as
 \begin{equation}
 \rho({ \vec r}, t) = \rho_{0}({ \vec r}) + \rho^{\prime} ({ \vec r}, t),
 \end{equation}
 where $\rho_{0}({\vec  r})$ is the mean density at position $  r$ and $\rho^{\prime}({\vec  r}, t)$ the fluctuation that depends both on the position and the time $t$. 
 The components of the moment of inertia become:
 \begin{equation}
 I_{ik} (t) = \int_{V} \rho({\vec  r}, t) x_{i}x_{k} \, dV =  \int_{V} [\rho_{0}({\vec  r}) + \rho^{\prime} ({ \vec r}, t)]  x_{i}x_{k} \, dV \equiv I_{ik}^{(0)} + I^{\prime}_{ik},
 \end{equation}
 where we have defined the time average of the component and its fluctuation. Because the average density is spherically symmetric, only the diagonal components of the $I^{(0)}_{ik}$ inertia tensor are different from zero. In other words 
 \begin{eqnarray}
 I_{xx}^{(0)} = I_{yy}^{(0)} = I_{zz}^{(0)} & =  & \frac{1}{3} \int_{V} \rho_{0}({ \vec r})\,  r^{2} \, dV,  \nonumber \\
 I_{ik}^{(0)} & = & 0 \mbox{ for $i\not= k$},  
 \end{eqnarray}
 and
 \begin{equation}
 {\rm Tr}\, I^{(0)} = 3I_{xx}^{(0)} =3 I_{yy}^{(0)} = 3 I_{zz}^{(0)}. 
 \end{equation}
 Therefore, only the fluctuating parts $I^{\prime}_{ik}$ contribute to the quadrupole moment tensor $Q_{ik}$. 
It is a symmetric tensor, therefore it is possible to reduce it to a diagonal form by a suitable orientation of the coordinate axes \citep[e.g.,][]{Goldstein50}.  Such axes are the principal axes of inertia of the star and their orientation varies randomly because of the density fluctuations. 

Because the trace of the quadrupole moment tensor is zero according to Eq.~(\ref{qdef}), only two of its diagonal components are independent, and we can write its non-vanishing components in the reference frame of the principal axes by introducing two scalars, $Q$ and $T$, as
\begin{equation}
\left\{
\begin{array}{l}
Q_{xx} \equiv Q + T/2 \\
Q_{yy} \equiv Q - T/2 \\
Q_{zz} \equiv - 2Q.  \\
\end{array} 
\right.
\label{qandt_def}
\end{equation}
Therefore, the gravitational potential becomes:
\begin{eqnarray}
\Phi & =  & -\frac{Gm_{\rm s}}{r} - \frac{3G}{2r^{3}} \frac{Q_{xx} x^{2} + Q_{yy} y^{2} + Q_{zz} z^{2}}{r^{2}}  = \nonumber \\
 & = & -\frac{Gm_{\rm s}}{r} - \frac{3G}{2r^{3}} Q - \frac{3G}{2r^{5}} \left[ \frac{1}{2} T \left( x^{2} - y^{2} \right) - 3Q z^{
2} \right],
\label{quad_pot_compl}
\end{eqnarray}
where we have separated the isotropic component of the quadrupole potential from the component depending on the coordinates $x=x_{1},y=x_{2},z=x_{3}$ of the point $P$ in the reference frame of the principal axes. We note that such coordinates continuously change in time because of the random re-orientation of the principal axes. Because the star is non-rotating, the statistical distribution of the orientation of the principal axes will be isotropic thanks to the spherical symmetry of the density fluctuations, the distribution of which will depend only on the distance $r$ from the center  of the star. 

In Sect.~\ref{power_sp}, we shall compute the autocorrelation function of the quadrupole potential and use it to compute the residual eccentricity of the orbit and the transit time variations. Such an autocorrelation cannot depend on the orientation of the principal axes because of the isotropy of the density fluctuations. Therefore, it cannot depend on the terms in square brackets in the right-hand side of Eq.~(\ref{quad_pot_compl}) because they depend on the coordinates of the point $P$ where the potential is evaluated in the reference frame of the principal axes\footnote{An alternative way to derive this result is to compute the autocorrelation of the quadrupole potential considering only the realizations in which the coordinates of the point $P$ in the reference frame of the principal axes are $(x,x,0)$ so that the term in square brackets in the right-hand side of Eq.~(\ref{quad_pot_compl}) vanishes. Such realizations are a subset of all possible realizations of our dynamical system, but such a subset is statistically equivalent to the whole ensemble of possible realizations of our system because of the  independence of its statistical properties on the orientation of the principal axes.}. In other words, we can limit ourselves to consider only the isotropic part of the quadrupole potential in our model which greatly simplifies our treatment 
\begin{equation}
\Phi = -\frac{Gm_{\rm s}}{r} - \frac{3G}{2r^{3}} Q. 
\label{gravit_poten}
\end{equation}
\vspace*{1mm}
%

\subsection{Perturbations of the planetary orbit}
\label{perturbations}
The fluctuating quadrupole gravitational potential induces perturbations of the orbital elements of the planet. Since such perturbations are small, we assume an unperturbed circular orbit for reference and use the Gauss equations to compute the time derivatives of the orbital elements \citep[e.g.,][\S~6.7.4]{Roy78}. From  Eq.~\eqref{gravit_poten}, it follows that the  perturbative acceleration acting on the planet is purely radial. We indicate such a radial acceleration with $S$, while the components of the perturbative acceleration in the orbital plane and normal to the radius vector, $T$, and normal to the orbital plane, $W$, are both zero:
\begin{eqnarray}
    S  =  -\frac{\partial \Phi}{\partial r} & = & -\frac{9G}{2 r^{4}} Q(t), \\
    T & = & 0, \\
    W & = & 0. 
\end{eqnarray}
The only two relevant Gauss equations for the variation in the orbital elements are:
\begin{eqnarray}
   \frac{de}{dt} &  = & \frac{S}{n a} \sin n t - \frac{e}{\tau_{\rm e}},  \label{dedt_gauss}\\
   \frac{d \epsilon}{dt} & = & -\frac{2 S}{na} \label{depsdt_gauss},
\end{eqnarray}
where $e$ is the eccentricity and $\epsilon$ the mean longitude at the epoch of the planetary orbit, $a$ its semimajor axis, $n$ the mean motion, $\tau_{\rm e}$ the tidal decay timescale of the eccentricity (cf. Eq.~\ref{tau_ecc}) and $t$ the time. Equations~\eqref{dedt_gauss} and~\eqref{depsdt_gauss} are written for an  unperturbed circular orbit with radius equal to $a$, while the action of the tides inside the star, that tends to damp the eccentricity, is represented by the term $-e/\tau_{\rm e}$. Such a term is not included into the Gauss equation for the eccentricity, but it is added here to take into account the effect of the stellar tides.


The mean longitude of the planet $l_{\rm p}$ is related to $\epsilon$ by \citep[cf.][]{Roy78}
\begin{equation}
    l_{\rm p} = \epsilon + n t,
\end{equation}
in the case of a circular reference orbit. Since $n$ can be assumed fixed on the timescales over which $\epsilon$ fluctuates (see below),  the fluctuations in the mean longitude are  $\Delta l_{\rm p} = \Delta \epsilon$, that can be used to compute the variations in the time of mid transit in the case of a planet that transits across the disc of our red giant star as
\begin{equation}
    O - C = \frac{\Delta \epsilon}{2\pi} P_{\rm orb}, 
    \label{ominusc_eq}
\end{equation}
where $O-C$ is the difference in the time of mid transit with respect to an unperturbed orbit of constant orbital period  $P_{\rm orb}=2\pi/n$.

\subsubsection{Residual eccentricity}
\label{eccentr_var}
We recast Eq.~\eqref{dedt_gauss} in the form
\begin{equation}
    \frac{de}{dt} + \frac{e}{\tau_{\rm e}(t)} = f(t), 
    \label{dedt_fo}
\end{equation}
where we made the dependence of the tidal damping timescale $\tau_{\rm e}$ on the time explicit (cf. Eq.~\ref{tau_ecc}) and defined
\begin{equation}
    f(t) \equiv -K Q(t) \sin nt,
    \label{def_ffunc}
\end{equation}
where 
\begin{equation}
K \equiv \frac{9G}{2 n a^{5}} = \frac{9n}{2 m_{\rm T} a^{2}}
\label{kappa_def}
\end{equation}
with the second expression for $K$ coming by applying Kepler third law. By applying the method of the variation of the constants, the general solution of Eq.~\eqref{dedt_fo} is
\begin{equation}
    e(t) = C_{0} \zeta(t) + \zeta(t) \int_{0}^{t} f(t^{\prime}) [\zeta(t^{\prime})]^{-1} \ d t^{\prime},
    \label{e_solut}
\end{equation}
where $C_{0}$ is a constant depending on the initial value of $e$ and 
\begin{equation}
    \zeta(t) \equiv \exp \left[ -\int_{0}^{t} \frac{dt^{\prime}}{\tau_{\rm e} (t^{\prime})} \right]. 
    \label{zeta_func}
\end{equation}
The ensemble mean value of the eccentricity $\langle e \rangle = 0$ because $f(t)$ is a stochastic function, while the mean value of its square can be obtained as
\begin{equation}
    \langle e^{2}(t) \rangle = \left\langle [\zeta(t)]^{2} \int_{0}^{t} \int_{0}^{t} f(t^{\prime}) f(t^{\prime\prime}) [\zeta(t^{\prime})]^{-1} [\zeta(t^{\prime\prime})]^{-1} \ dt^{\prime} dt^{\prime\prime} \right\rangle.
\label{e2_aver_two}
\end{equation}
The term containing the constant factor $C_{0}$ in Eq.~\eqref{e_solut} becomes negligible after a sufficiently long interval of time because it decreases exponentially according to Eq.~\eqref{zeta_func}. For this reason, the terms containing $C_{0}$ have been dropped from Eq.~\eqref{e2_aver_two}. 
The ensemble mean commutes with the integration allowing us to evaluate $\langle e^{2}(t) \rangle$ if we known $\langle f(t^{\prime}) f(t^{\prime\prime}) \rangle$. 

The characteristic timescale for the evolution of the residual eccentricity is on the order of $\tau_{\rm e}$, that is, much longer than the correlation timescale of the fluctuations of the quadrupole moment $Q(t)$ that determine the function $f(t)$ because the quadrupole fluctuations occur on the timescale of the convective motions.  In other words, we can regard $f(t)$ as a delta-correlated process, that is,
\begin{equation}
    \left\langle f(t^{\prime}) f(t^{\prime\prime}) \right\rangle = D\, \delta (t^{\prime\prime} - t^{\prime}), 
    \label{delta_corr_proc}
\end{equation}
where $D$ is a slowly varying function of the time and $\delta$ is the Dirac delta function. By substituting Eq.~\eqref{delta_corr_proc} into Eq.~\eqref{e2_aver_two}, we find
\begin{eqnarray}
    \langle e^{2}(t) \rangle   = [\zeta(t)]^{2} \int_{0}^{t} \int_{0}^{t} D\, \delta (t^{\prime\prime} - t^{\prime}) [\zeta(t^{\prime})]^{-1} [\zeta(t^{\prime\prime})]^{-1} \ dt^{\prime} dt^{\prime\prime} \\
    = [\zeta(t)]^{2} \int_{0}^{t} D(t^{\prime}) [ \zeta(t^{\prime})]^{-2}\, d t^{\prime},
    \label{e2_integr}
\end{eqnarray}
where we have explicitly introduced the slow time dependence of $D$ that is relevant on the timescales over which $\langle e^{2} \rangle$ evolves. 

We can evaluate $D(t)$ considering a time interval much longer than the correlation timescale of the stochastic function $f(t)$, yet short enough that the parameters $a$ and $n$ of the unperturbed orbit and the mass of the star, $m_{\rm s}$ are almost constant, so that $K$, $\tau_{\rm e}$, and the slowly varying $D(t)$ can be regarded as constant. In this case, $\langle e^{2} \rangle$ can be obtained by a simple integration from Eq.~\eqref{e2_integr} as 
\begin{equation}
    \langle e^{2} \rangle = \frac{1}{2} D \, \tau_{\rm e},
    \label{e2_D}
\end{equation}
for $t \gg \tau_{\rm e}/2$. On the other hand, $\langle e^{2} \rangle$ can be obtained from the power spectrum of $e$ that can be computed by taking the Fourier transform of Eq.~\eqref{dedt_fo}. 

We introduce the Fourier transform of a function $g(t)$ of the time $t$ as
\begin{equation}
\tilde{g} (\omega) \equiv  \int_{-\infty}^{\infty} g(t) \exp(-i\omega t)\, dt,
\label{fourier_transf_def}
\end{equation}
where $\omega$ is the frequency and $i = \sqrt{-1}$ the imaginary unit. The inverse transform is:
\begin{equation}
g(t)  = \frac{1}{2\pi} \int_{-\infty}^{\infty} \tilde{g}(\omega ) \exp(i\omega t)\, d\omega. 
\end{equation}
Taking the Fourier transform of both sides of Eq.~\eqref{dedt_fo}, we have
\begin{equation}
    \tilde{e} (\omega) = \frac{\tilde{f}(\omega) \left({\tau_{\rm e}^{-1}} - i\omega \right)}{\tau_{\rm e}^{-2} + \omega^{2}}. 
    \label{e_fouriert}
\end{equation}
The power spectrum of $e$ is the Fourier transform of its autocorrelation function $R_{\rm e}(\tau) \equiv \langle e(t+\tau) e(t) \rangle$, where $\tau$ is a time lag, so that 
\begin{equation}
    \langle e^{2} \rangle = R_{\rm e}(0) = \frac{1}{2\pi} \int_{-\infty}^{\infty} P_{\rm e}(\omega) \ d\omega,
    \label{psd}
\end{equation}
where $P_{\rm e} (\omega) = \tilde{e}(\omega) \tilde{e}^{*}(\omega)$ is the power spectrum of the eccentricity and the asterisk indicates complex conjugation. Making use of Eq.~\eqref{e_fouriert}, Eq.~\eqref{psd} becomes
\begin{equation}
    \langle e^{2} \rangle = \frac{1}{2\pi} \int_{-\infty}^{\infty} \frac{P_{\rm f} (\omega)}{{\tau_{\rm e}^{-2}} + \omega^{2}} \ d\omega,
    \label{e2_psd_int}
\end{equation}
where $P_{\rm f}(\omega) = \tilde{f}(\omega) \tilde{f}^{*}(\omega)$ is the power spectrum of the stochastic function $f(t)$. Since the tidal damping timescale $\tau_{\rm e}$ is very long, the integrand in Eq.~\eqref{e2_psd_int} is very large   for $\omega=0$ and decreases sharply when $|\omega | > 0$. In other words, it can be well approximated as
\begin{equation}
    \langle e^{2} \rangle \simeq \frac{P_{\rm f}(0)}{2\pi} \int_{-\infty}^{\infty} \frac{d\omega}{{\tau_{\rm e}^{-2}} + \omega^{2}} = \frac{1}{2} P_{\rm f}(0) \, \tau_{\rm e},
    \label{e2_psd_int1}
\end{equation}
To evaluate $P_{\rm f}(0)$, we use Eq.~\eqref{def_ffunc}, the Euler formula for the sine, and Eq.~\eqref{fourier_transf_def} to compute 
\begin{eqnarray}
    \tilde{f}(0) = \int_{-\infty}^{\infty} -K Q(t) \frac{1}{2i} \left[ \exp(int)-\exp(-int) \right] \, dt \\
    = \frac{K}{2i}\left[ \tilde{Q}(n) - \tilde{Q}(-n) \right] = K \Im \left\{\tilde{Q}(n) \right\},
\end{eqnarray}
where $\tilde{Q}$ is the Fourier transform of $Q(t)$ that, being a real function, verifies $\tilde{Q}^{*}(n) = \tilde{Q}(-n)$. Given that $Q(t)$ is a stochastic variable, $[\Re \tilde{Q}(\omega)]^{2} = [\Im \tilde{Q}(\omega)]^{2} = P_{\rm Q} (\omega)/2$ in a statistical sense, where $P_{\rm Q}(\omega)$ is the power spectrum of the fluctuating quadrupole moment $Q(t)$. By applying this result, we have 
\begin{eqnarray}
    \langle e^{2} \rangle = \frac{1}{4} K^{2} \tau_{\rm e} P_{\rm Q} (n)
    = \frac{81}{16} \frac{n^{2} \tau_{\rm e} P_{\rm Q}(n)}{m_{\rm T}^{2} a^{4}},
    \label{e2_psd_fin}
\end{eqnarray}
that can be compared with Eq.~\eqref{e2_D} to find
\begin{equation}
    D(t) =  \frac{81}{8} \frac{n^{2} P_{\rm Q}(n)}{m_{\rm T}^{2} a^{4}}.
    \label{D_eq}
\end{equation}
Substituting Eq.~\eqref{D_eq} into Eq.~\eqref{e2_integr}, we  find the equation for the evolution of the expectation value of the square of the residual  eccentricity as
\begin{equation}
    \langle e^{2}(t) \rangle = \frac{81}{8} [\zeta(t)]^{2}\int_{0}^{t} \frac{n^{2} P_{\rm Q}(n)}{m_{\rm T}^{2} a^{4}} [\zeta(t^{\prime})]^{-2} \, dt^{\prime}, 
    \label{residual_e2}
\end{equation}
where we have explicitly indicated the time dependence only for the function $\zeta$ given by Eq.~\eqref{zeta_func}, even if all the quantities appearing in the integrand are functions of the time. An alternative way to evaluate $\langle e^{2} \rangle$, according to the method proposed by \citet{Phinney92}, is presented in Appendix~\ref{appendix_B}. 

{  The residual eccentricity $\langle e^{2} \rangle$ is practically independent of the mass of the planet because $m_{\rm p} \ll m_{\rm s} \simeq m_{\rm T}$ and it enters into Eq.~\eqref{residual_e2} only through the function $\zeta(t)$ via the timescale $\tau_{\rm e}$ (cf. Eq.~\ref{zeta_func}). Considering the expression of $\tau_{\rm e}$ given by Eq.~\eqref{tau_ecc}, the factors with the planet mass cancel out in Eq.~\eqref{residual_e2} because $m_{\rm p}$ is a constant that can be taken out of the integration. As a matter of fact, the independence of the residual eccentricity on the planet mass is true also for the more general equation~\eqref{e2_aver_two}, provided that $f(t)$ depends only on stellar quantities. }

The statistical distribution of the residual eccentricity can be computed from the theory of stochastic processes  \citep[e.g.,][]{Phinney92,LanzaRodono01}. Nevertheless, a  straightforward derivation exploits a method used to compute the Maxwell-Boltzmann distribution of molecular velocities in an ideal gas as illustrated in Appendix~\ref{appendix_A}. Indicating with $p_{\rm e} (e) \, de$ the probability that the residual eccentricity falls into the interval $[e, e+de]$, the result is 
\begin{equation}
p_{\rm e}(e) = \sqrt{\frac{2}{\pi \langle e^{2} \rangle}} \exp \left( - \frac{e^{2}}{2\langle e^{2} \rangle}\right). 
\label{eres_distr}
\end{equation}
Therefore, the probability $P(e \leq e_{0})$ that the residual eccentricity $e$ be smaller than or equal to a given value $e_{0}$ can be expressed in terms of the error function as
\begin{equation}
P(e \leq e_{0}) = {\rm erf} \left( \frac{e_{0}}{\sqrt{2 \langle e^{2} \rangle}} \right), \label{pe_cumul_prob}
\end{equation}
where we define the error function as
\begin{equation}
{\rm erf}(z) \equiv \frac{2}{\sqrt{\pi}} \int_{0}^{z} \exp(-\xi^{2}) \, d\xi,  
\label{probelte0}
\end{equation}
so that $\lim_{z\rightarrow \infty} {\rm erf}(z) =1$. 

\subsubsection{Mean longitude at the epoch}
\label{mean_long_var}

The formal solution of Eq.~\eqref{depsdt_gauss} is
\begin{equation}
    \epsilon(t) = -2K \int_{0}^{t} Q(t^{\prime}) \, dt^{\prime},
\end{equation}
assuming the initial condition $\epsilon(0)=0$. Since the quadrupole moment $Q$ is a stochastic variable with $\langle Q \rangle = 0$ also $\langle \epsilon \rangle =0$, but $\langle \epsilon^2 \rangle$ is different from zero and is a function of the time $t$ because $\epsilon$ performs a Brownian motion around the point $\epsilon = 0$. Specifically, the ensemble mean of $\epsilon^{2}$ is 
\begin{equation}
    \langle \epsilon^{2} \rangle = 4 K^{2} \int_{0}^{t} \int_{0}^{t} \langle Q(t^{\prime}) Q(t^{\prime\prime}) \rangle dt^{\prime} dt^{\prime\prime}. 
\end{equation}
Considering time intervals much longer than the correlation time of the quadrupole moment fluctuations, we can assume a delta-correlated process with $\langle Q(t^{\prime}) Q(t^{\prime\prime}) \rangle = E \delta (t^{\prime\prime} - t^{\prime})$ finding $\langle \epsilon^{2} \rangle = 4K^{2} E t$ as expected in the case of a Brownian motion. 

To find the value of $E$, we use an approach based on the theory of the Brownian motion and consider that $Q$ at any given time is the sum of the contributions   coming from the different layers inside the convection zone of the star. Such contributions are uncorrelated with each other, while each of them has an autocorrelation timescale equal to the convective turnover time $\tau_{\rm c}(r)$ at the radius $r$ of the layer itself. We indicate the layers inside the stellar convection zone with an index $j$ and assume that the layer centered at radius $r_{j}$ contributes a variation $\Delta \epsilon_{j}$ given by
\begin{equation}
    \Delta \epsilon_{j} \sim Q_{j}(t) \tau_{\rm c}(r_{j}),
    \label{delta_epsk}
\end{equation}
where $Q_{j}(t)$ is the contribution to the quadrupole moment fluctuation at the time $t$ coming from the $j$-th layer. We introduce the probability density function $p_{j}(\Delta \epsilon_{j}, t)$ giving the probability $p_{j}(\Delta \epsilon_{j}, t)\, d(\Delta \epsilon_{j})$ of having a fluctuation in the interval $[\Delta \epsilon_{j}, \Delta \epsilon_{j} + d(\Delta \epsilon_{j})]$ at the time $t$. By definition
\begin{equation}
    \int_{-\infty}^{\infty} p_{j} (\Delta \epsilon_{j}, t) d (\Delta \epsilon_{j})  = 1. 
    \label{prob_dens_norm}
\end{equation}
The probability ${\cal P}_{j}$ of having a fluctuation $ \epsilon_{j}$ at the time $t+dt$ as a consequence of the fluctuations of the quadrupole moment contribution of the $j$-th layer is 
\begin{eqnarray}
    {\cal P}_{j} (\epsilon_{j}, t+ dt) = \int_{-\infty}^{\infty} {\cal P}_{j}(\epsilon_{j} - \Delta \epsilon_{j}, t) p_{j} (\Delta \epsilon_{j}, t)  \, d (\Delta \epsilon_{j}) =  \\
    \int_{-\infty}^{\infty} \left[ {\cal P}_{j}(\epsilon_{j}, t) p_{j}(\Delta \epsilon_{j}, t) -  \frac{\partial {\cal P}_{j}}{\partial \epsilon_{j}} \Delta \epsilon_{j} p_{j} (\Delta \epsilon_{j}, t) + \right. \nonumber \\ 
   \left.  \frac{1}{2} \frac{\partial^{2} {\cal P}_{j}}{\partial \epsilon_{j}^{2}} (\Delta \epsilon_{j})^{2} p_{j}(\Delta \epsilon_{j}, t) + ...  \right] \, d (\Delta \epsilon_{j}),
   \label{eq50}
\end{eqnarray}
where we developed the integrand function ${\cal P}_{j}$ in a Taylor series retaining only the terms up to the second order and considered that $\Delta \epsilon_{j}$ is the variation in $\epsilon_{j}$ occurring along the small time step $dt$, whose probability distribution is given by $p_{j}(\Delta \epsilon_{j}, t)$ at the time $t$. 

The integral of the first term in square brackets on the right-hand side of Eq.~\eqref{eq50} is ${\cal P}_{j} (\epsilon_{j},t)$ thanks to the normalization of the probability density $p_{j}(\Delta \epsilon_{j}, t)$ given by Eq.~\eqref{prob_dens_norm}, while the integral of the second term vanishes because $p_{j}(\Delta \epsilon_{j}, t) = p_{j} (-\Delta \epsilon_{j}, t)$. On the other hand, we can develop ${\cal P}_{j}$ in a time series and retain only the first-order term as 
\begin{equation}
    {\cal P}_{j} (\epsilon_{j}, t+dt) = {\cal P}_{j}(\epsilon_{j}, t) + \frac{\partial {\cal P}_{j}}{\partial t} \tau_{c}(r_{j}) + ..., 
    \label{eq51}
\end{equation}
where we assumed the small time increment $dt$ equal to the autocorrelation time of the local quadrupole moment fluctuations, $\tau_{\rm c}(r_{j})$,  in order to apply Eq.~\eqref{delta_epsk} in the later developments. 
By comparing Eqs.~\eqref{eq50} and~\eqref{eq51} and taking into account Eqs.~\eqref{delta_epsk} and~\eqref{prob_dens_norm}, we find
\begin{equation}
    \frac{\partial {\cal P}_{j}}{\partial t} = {\cal D}_{j} \frac{\partial^{2} {\cal P}_{j}}{\partial \epsilon_{j}^{2}},
    \label{diff_eq}
\end{equation}
with  
\begin{equation}
    {\cal D}_{j} \equiv \frac{1}{2} \int_{-\infty}^{\infty} 4 K^{2} Q_{j}^{2} \tau_{\rm c} (r_{j}) p_{j} (Q_{j}, t) \, dQ_{j} = 2 K^{2} \langle Q_{j}^{2} \rangle \tau_{\rm c} (r_{j}), 
    \label{D_diff_eq}
\end{equation}
where we made use of Eq.~\eqref{delta_epsk} to transform the integration variable $\Delta \epsilon_{j}$ into the corresponding quadrupole moment fluctuation $Q_{j}$ whose probability density function is $p_{j} (Q_{j}, t)$. Equation~\eqref{diff_eq} is a diffusion equation whose solution is
\begin{equation}
    {\cal P}_{j} (\epsilon_{j}, t) = \frac{1}{\sqrt{2\pi {\cal D}_{j} t}} \exp \left( -\frac{\epsilon_{j}^{2}}{2{\cal D}_{j} t}\right).
\end{equation}
Therefore, the variance of $\epsilon_{j}$ is $\langle \epsilon_{j}^{2} \rangle = {\cal D}_{j} t$ and increases linearly in time as expected in the case of a system performing a Brownian motion. The variance of $\epsilon$ can be found by summing all the uncorrelated contributions coming from the single layers inside the stellar convection zone. This yields
\begin{equation}
    \langle \epsilon^{2} \rangle = \sum_{j} \langle \epsilon_{j}^{2} \rangle = 2K^{2} \, t \, \int_{r_{\rm b}}^{R_{\rm s}} \langle Q^{2} (r) \rangle \tau_{\rm c} (r) dr,
    \label{epsilon2_eq}
\end{equation}
where $r_{\rm b}$ is the radius at the base of the stellar convection zone and $R_{\rm s}$ the stellar radius, and we have dropped the index $j$ specifying the layer inside the star because the sum over the layers has been substituted by the integration. We do not consider the time dependence of $K$ and $\langle Q^{2}(r) \rangle$ in Eq.~\eqref{epsilon2_eq} because it is relevant only on evolutionary timescales, that is, on time intervals much longer than any observational interval over which $\langle \epsilon^{2} \rangle$ can be measured.

The fluctuations in the time of mid transit can be derived from Eq.~\eqref{ominusc_eq} and are:
\begin{eqnarray}
\langle O - C \rangle & = & 0, \label{o-c1}\\
\langle (O-C)^{2} \rangle &  = & \frac{\langle \epsilon^{2} \rangle}{2\pi} P_{\rm orb}, \label{o-c2}
\end{eqnarray}
where Eq.~\eqref{o-c1} follows from $\langle \epsilon \rangle=0$, while $\langle \epsilon^{2} \rangle$ in Eq.~\eqref{o-c2} is given by Eq.~\eqref{epsilon2_eq}. Therefore, $O-C$  makes a Brownian motion around zero with a standard deviation that increases linearly in time. In principle, this can be used to measure it, provided that we have observations extended over a sufficiently long interval of time as we shall consider in Sect.~\ref{applications}. 

\subsection{Power spectrum of the quadrupole moment fluctuations}
 \label{power_sp}
 
To compute the residual eccentricity according to Eq.~(\ref{residual_e2}), we need to determine the power spectrum of the time-dependent quadrupole moment $Q(t)$. From the power spectrum of the local contribution to the quadrupole moment fluctuations at radius $r$, the mean squared value $\langle Q^{2}(r) \rangle$ appearing in  Eq.~\eqref{epsilon2_eq} follows, thus allowing us to compute the variance of the mean longitude at the epoch. 

First we consider the power spectrum of the total quadrupole moment of the star,  $Q(t)$, and make use of the property that the power spectrum is the Fourier transform of the autocorrelation function $R_{Q}(\tau)$ of  $Q(t)$. The autocorrelation function is the expectation value 
 $R_{Q} (\tau) \equiv \langle Q(t+\tau) Q (t) \rangle$, where $\tau$ is a time lag, and we make use of the definition (\ref{fourier_transf_def}) of the Fourier transform to obtain 
  \begin{equation}
 P_{Q} (\omega) = \int_{-\infty}^{\infty} R_{Q} (\tau) \exp(-i\omega \tau) \, d\tau. 
 \end{equation}
 The quadrupole $Q$ can be expressed as (cf. Eq.~\ref{qandt_def})
 \begin{equation}
 Q = -\frac{1}{2} Q_{zz} = -\frac{1}{2} (I_{zz} - \frac{1}{3} {\rm Tr} I) = -\frac{1}{3} I^{\prime}_{zz} + \frac{1}{6} (I^{\prime}_{xx} + I^{\prime}_{yy} ). 
 \end{equation}
 The fluctuations $I_{xx}^{\prime}$, $I_{yy}^{\prime}$, and $I_{zz}^{\prime}$ are uncorrelated with each other and have the same autocorrelation function because of the spherical symmetry of the turbulent velocity field, that is, $R_{I^{\prime}_{xx}} (\tau) = R_{I^{\prime}_{yy}} (\tau)= R_{I^{\prime}_{zz}} (\tau) $.  This implies that 
 \begin{equation}
 R_{Q} (\tau) = \frac{1}{6} \, R_{I_{xx}^{\prime}} (\tau). 
 \label{auto_qp_ip}
 \end{equation}
Therefore, we are left with the computation of the autocorrelation function of the fluctuations of the principal moment of inertia $I^{\prime}_{xx}$. Considering the isotropy of the fluctuations in our model
\begin{equation}
I^{\prime}_{xx} = \frac{1}{3} \left( I^{\prime}_{xx} + I^{\prime}_{yy} + I^{\prime}_{zz} \right) = \frac{1}{3} \int_{V} \rho^{\prime} ({ \vec r}, t) \, r^{2} \, dV, 
\end{equation}
where ${  r}$ is the position vector, the modulus $r$ of which is the radial distance from the centre of the star. 
The autocorrelation of $I^{\prime}_{xx}$ becomes 
\begin{multline}
R_{I^{\prime}_{xx}} (\tau)  =  \langle I^{\prime}_{xx} (t+\tau)\, I_{xx}^{\prime} (t) \rangle = \\ 
= \frac{1}{9} \langle \int_{V} \rho^{\prime} ({ \vec r}, t+\tau)\, r^{2} \, dV \times \int_{V} \rho^{\prime} ({ \vec r}^{\prime}, t)\, r^{\prime \, 2} \, dV^{\prime} \rangle = \\
= \frac{1}{9} \int_{V} \int_{V} \langle \rho^{\prime} ({ \vec r}, t+\tau) \rho^{\prime} ({ \vec r}^{\prime}, t) \rangle \, r^{2} r^{\prime \, 2}\, dV \, dV^{\prime}. 
\end{multline}
The autocorrelation function of the density fluctuations can be written as the product of a spatial autocorrelation and a temporal autocorrelation. In the framework of the mixing-length theory, that is, a local theory of convection,  the density fluctuations at a given radius $r$ are spatially correlated over a volume of order $\ell^{3}(r)$, where $\ell(r)$ is the mixing length at radius $r$, while their correlation time is $\tau_{\rm c} (r) = \ell(r)/\varv_{\rm c}(r)$, where $\varv_{\rm c}(r)$ is the convective velocity at radius $r$ as provided by the mixing-length (or any more sophisticated) convection model. 

If we make the hypothesis that $\ell \ll r$ in the convective envelope and that the fluctuations outside each convective cell of volume $\ell^3$ are uncorrelated with each other, we can simply integrate all the local autocorrelations of the density fluctuations and obtain\footnote{In the case of a very small mixing length ($\ell \ll r$), one could write the autocorrelation function by introducing a Dirac delta function of the spatial coordinates as $\langle \rho^{\prime} ({ \vec r}, t+\tau) \rho^{\prime} ({\vec   r}^{\prime}, t) \rangle = \ell^{3}(r) \delta ({ \vec r}^{\prime} - { \vec r}) \exp[-|\tau| / \tau_{\rm c}(r)]$. In this way, the integration over $dV^{\prime}$ is immediately performed and the double integral  becomes a simple integral over $dr$. }
\begin{eqnarray}
R_{I^{\prime}_{xx}} (\tau)  \simeq   \frac{1}{9} \int_{V} [\rho^{\prime}({ \vec r}, 0)]^{2} \ell^{3}(r)\, r^{4} \exp[-|\tau|/ \tau_{c}(r)] \, dV= \nonumber \\
= \frac{4\pi}{9} \int_{r_{\rm b}}^{R_{\rm s}} [\rho^{\prime}(r)]^{2} \ell^{3}(r)\, r^{6} \exp[-|\tau|/ \tau_{c}(r)] \, dr,   
\label{autoc_ixx}
\end{eqnarray}
where we assume that the local density fluctuations are isotropic on the average and do not explicitly depend on the time, $r_{\rm b}$ is the radius at the lower boundary of the stellar convection zone, and $R_{\rm s}$ is the radius of the star. 

The local density fluctuations can be estimated by means of the mixing-length theory \citep[see Ch.~7 of][]{KWW12} by equating the average work per unit volume done by the buoyancy force  along the path of a convective element (equal to the mixing length $\ell(r)$) to the kinetic energy density in the convective motions 
\begin{equation}
\frac{1}{2} \rho_{0} (r) \varv_{\rm c}^{2} (r) \sim \frac{1}{2} g(r) |\rho^{\prime} (r) | \ell (r), 
\label{dens_fluc}
\end{equation}
where $g(r)= G m(r) r^{-2}$ is the acceleration of gravity with $m(r)$ the mass of the star inside the radius $r$. 
Therefore, we find
\begin{equation}
 |\rho^{\prime} (r)| \sim \frac{\rho_{0}(r) \varv_{\rm c}^{2}(r)}{g(r) \ell(r)}.
 \label{conv_fluc}
\end{equation}
Substituting Eq.~(\ref{conv_fluc}) into Eq.~(\ref{autoc_ixx}), we find
\begin{equation}
R_{I^{\prime}_{xx}} (\tau)  \simeq \frac{4\pi}{9}  \int_{r_{\rm b}}^{R_{\rm s}} \left[ \frac{\rho_{0}(r) \varv_{\rm c}^{2}(r)}{g(r) \ell(r)} \right]^{2} \ell^{3}(r) \, r^{6} \exp[-|\tau|/ \tau_{c}(r)] \, dr,
\label{autoc_Ipxx}
\end{equation}
where the quantities $\rho_{0}(r)$, $\varv_{\rm c}(r)$, $\ell(r)$, $\tau_{\rm c}(r)$, and $g(r)$ follow from a model of the stellar structure at the given time $t$. 

To facilitate the computation of the power spectrum of $I^{\prime}_{xx}$, we compute the Fourier transform of the exponential decorrelation function of the time that appears in Eq.~(\ref{autoc_Ipxx}):
\begin{equation}
E_{\rm D}(\tau) \equiv \exp(-|\tau| / \tau_{0}).
\end{equation}
We have 
\begin{multline}
\tilde{E}_{\rm D}(\omega) \equiv  \int_{-\infty}^{\infty} \exp (-|\tau| / \tau_{0}) \exp(-i\omega \tau) \, d\tau =  \\
= \int_{-\infty}^{0} \exp (\tau / \tau_{0}) \exp(-i\omega \tau) \, d\tau \, + \int_{0}^{\infty} \exp (-\tau / \tau_{0}) \exp(-i\omega \tau) \, d\tau =  \\
= \int_{-\infty}^{0} \exp \left[ \left( \frac{1}{\tau_{0}} - i\omega \right) \tau \right] \, d\tau \, + \int_{0}^{\infty} \exp \left[ - \left( \frac{1}{\tau_{0}} + i\omega \right) \tau \right] \, d\tau =  \\
= \frac{1}{\left({\tau_{0}^{-1}} - i\omega \right)} + \frac{1}{\left( {\tau_{0}^{-1}} + i\omega \right)} = \frac{2{\tau_{0}^{-1}}}{{\tau_{0}^{-2}} + \omega^{2}} = \frac{2\tau_{0}}{1+\tau_{0}^{2} \omega^
{2}}. 
\label{e_decorr}
\end{multline}
The power spectrum of $I^{\prime}_{xx} (t)$ follows from the Fourier transform of Eq.~(\ref{autoc_Ipxx}). It can be immediately computed because the Fourier transform implies an integral over $\tau$ that can be taken out of the integral over the radius $r$ giving, thanks to Eq.~(\ref{e_decorr}),
\begin{equation}
P_{I^{\prime}_{xx}} (\omega) \sim \frac{8\pi}{9}  \int_{r_{\rm b}}^{R_{\rm s}} \left[ \frac{\rho_{0}(r) \varv_{\rm c}^{2}(r)}{g(r) \ell(r)} \right]^{2} \ell^{3}(r) \, r^{6} \frac{\tau_{\rm c}(r)}{\tau_{\rm c}^{2}(r)\, \omega^{2}+1} \, dr. 
\label{eq67}
\end{equation}
The power spectrum of $Q$ follows from the Fourier transform of both sides of Eq.~(\ref{auto_qp_ip}), thus giving 
\begin{equation}
P_{Q} (\omega) = \frac{1}{6} P_{I^{\prime}_{xx}} (\omega). 
\label{eq63}
\end{equation}
Specifically, $P_{Q}(n)$ appearing into Eq.~(\ref{residual_e2}), is given by 
\begin{equation}
P_{Q} (n) \sim \frac{4\pi}{27}  \int_{r_{\rm b}}^{R} \left[ \frac{\rho_{0}(r) \varv_{\rm c}^{2}(r)}{g(r) \ell(r)} \right]^{2} \ell^{3}(r) \, r^{6} \frac{\tau_{\rm c}(r)}{\tau_{\rm c}^{2}(r)\, n^{2}+1} \, dr. 
\label{power_q(n)}
\end{equation}
To evaluate the integral appearing in Eq.~\eqref{epsilon2_eq}, we compute the mean variance of the local contribution to the quadrupole moment fluctuations at radius $r$ from Eq.~\eqref{eq63} and the integrand in Eq.~\eqref{eq67}, that is, the power spectrum of the local fluctuations of the moment of inertia. We have
\begin{equation}
    \langle Q^{2}(r) \rangle \, \tau_{c}(r) \rangle = \frac{2}{27} \left[ \frac{\rho_{0}(r) \varv_{c}^{2}(r)}{g(r) \ell(r)}\right]^{2} \ell^{3}(r) \, r^{6} \int_{-\infty}^{\infty} \frac{\tau_{\rm c}^{2}(r)}{\tau_{\rm c}^{2}(r)\,  \omega^{2} +1} \, d\omega,
\end{equation}
that can be substituted into Eq.~\eqref{epsilon2_eq} after computing the integral over $\omega$ to give
\begin{equation}
    \langle \epsilon^{2} \rangle  = \frac{3\pi \, n^{2}}{m_{\rm T}^{2}\, a^{4}} \left\{ \int_{r_{\rm b}}^{R_{\rm s}} \left[ \frac{\rho_{0}(r) \varv_{c}^{2}(r)}{g(r) \ell(r)}\right]^{2} \ell^{3}(r) \, \tau_{\rm c} (r)\, r^{6}\, dr, \right\} t 
    \label{epsilon_o-c2}
\end{equation}
where we substituted for $K$ from Eq.~\eqref{kappa_def}.  

\section{Stellar structure and evolution}
\label{stellar_str_evol_model}
To apply the above model to an Earth-like planet orbiting a Sun-like star, we first introduce a model for the internal structure and evolution of the host star. 
We used the r21.12.1 release of the MESA stellar evolution code 	\citep[][]{2011ApJS..192....3P, 2013ApJS..208....4P, 2015ApJS..220...15P, 2018ApJS..234...34P, 2019ApJS..243...10P}  to calculate the structure and evolution of a one solar mass star from the pre-main-sequence phase to the tip of the asymptotic giant branch. 
	 
We considered the standard inputs and recipes recommended by the MESA team for the equation of state, nuclear reaction rates and screening factors  \citep{2019ApJS..243...10P}, and for the radiative and conductive opacities \citep{2011ApJS..192....3P}.  As for the mixing-length theory of convection, we adopted the formulation by  \citet{1965ApJ...142..841H}. The Ledoux's \citeyearpar{1947ApJ...105..305L} criterion was used to assess convective stability. 
	
The occurrence of core overshooting during the core helium burning phase (CHeB) has been suggested by \citet{2015MNRAS.453.2290B} to satisfy both asteroseismic constraints on the CHeB phase and the observed luminosity at the AGB bump. 
 Then, considering observed constraints on the AGB bump of solar mass stars of solar metallicity, \citet{2022A&A...668A.115D} estimated that the ratio of the overshooting extent to the pressure scale height is in the range $\alpha_\mathrm{ov,CHeB}= 0.25-0.50$, following a step scheme \citep{1975A&A....40..303M}. In our model, we therefore took into account the moderate amount of convective core overshooting, $\alpha_\mathrm{ov,CHeB}= 0.25$. The temperature gradient in the overshooting region is taken to be radiative. During CHeB, the convective core grows with time, the convective boundaries must therefore be correctly located to allow the emergence of semi-convective regions, which is delicate. To cope with these difficulties, we used the new Convective Pre-Mixing (CPM) scheme recommended by the MESA team  \citep{2019ApJS..243...10P}. Furthermore, we considered step overshooting from the bottom of the convective envelope, with a value $\alpha_\mathrm{ov, env}= 0.3$. This value has been obtained by \citet{2018ApJ...859..156K} on the basis of constraints on the position of the RGB bump of $\sim3000$ stars.  Moreover, concerning mixing, we did not take into account microscopic diffusion, nor rotational mixing, nor thermohaline mixing.
	
We took mass loss into account both on the RGB and on the AGB. From the base of the RGB up to the end of the core helium burning phase,  we used Reimers's prescription \citeyearpar{1975MSRSL...8..369R} with the moderate, maximum value $\eta_R=0.3$ of the mass loss parameter, inferred by \citet{2012MNRAS.419.2077M,2021A&A...645A..85M}  from asteroseismic constraints on red giants in the Kepler field. On the AGB, we used Blöcker's \citeyearpar{1995A&A...297..727B} prescription with $\eta_B=0.02$ as prescribed by \citet{2020A&A...641A.103V}.
    As for the atmosphere boundary condition, we used an Eddington's grey $T(\tau)$ relation with $\tau$ being here the optical depth.
	
We adopted the GS98 solar mixture \citep{1998SSRv...85..161G} as a reference. It corresponds to a present solar photospheric metal $Z$ to hydrogen $X$ mass fraction ratio $(Z/X)_\odot=0.0229$. The solar model calibration, that is, the requirement that, at the solar age, the solar model reaches the solar radius and luminosity\footnote{We took the IAU 2015 resolution B3  values for the solar mass $M_\odot=1.9884\,\times 10^{30}$\, kg, radius $R_\odot=6.957\, \times 10^{8}$\, m, and luminosity $L_\odot=3.828\,\times 10^{26}\, \mathrm{W}$. The solar age is fixed to $A_\sun=4.57\, \times 10^9\, \mathrm{yr}$ \citep{2007leas.book...45C}.}, has provided the solar initial helium abundance in mass fraction  $Y_0=0.2602$ and a mixing-length parameter value of  $\alpha_\mathrm{MLT, \odot}=1.950$.

\begin{figure}
\vspace*{-25mm}
\hspace*{-10mm} 
 \centering{
\includegraphics[width=11cm,height=20cm,angle=0]{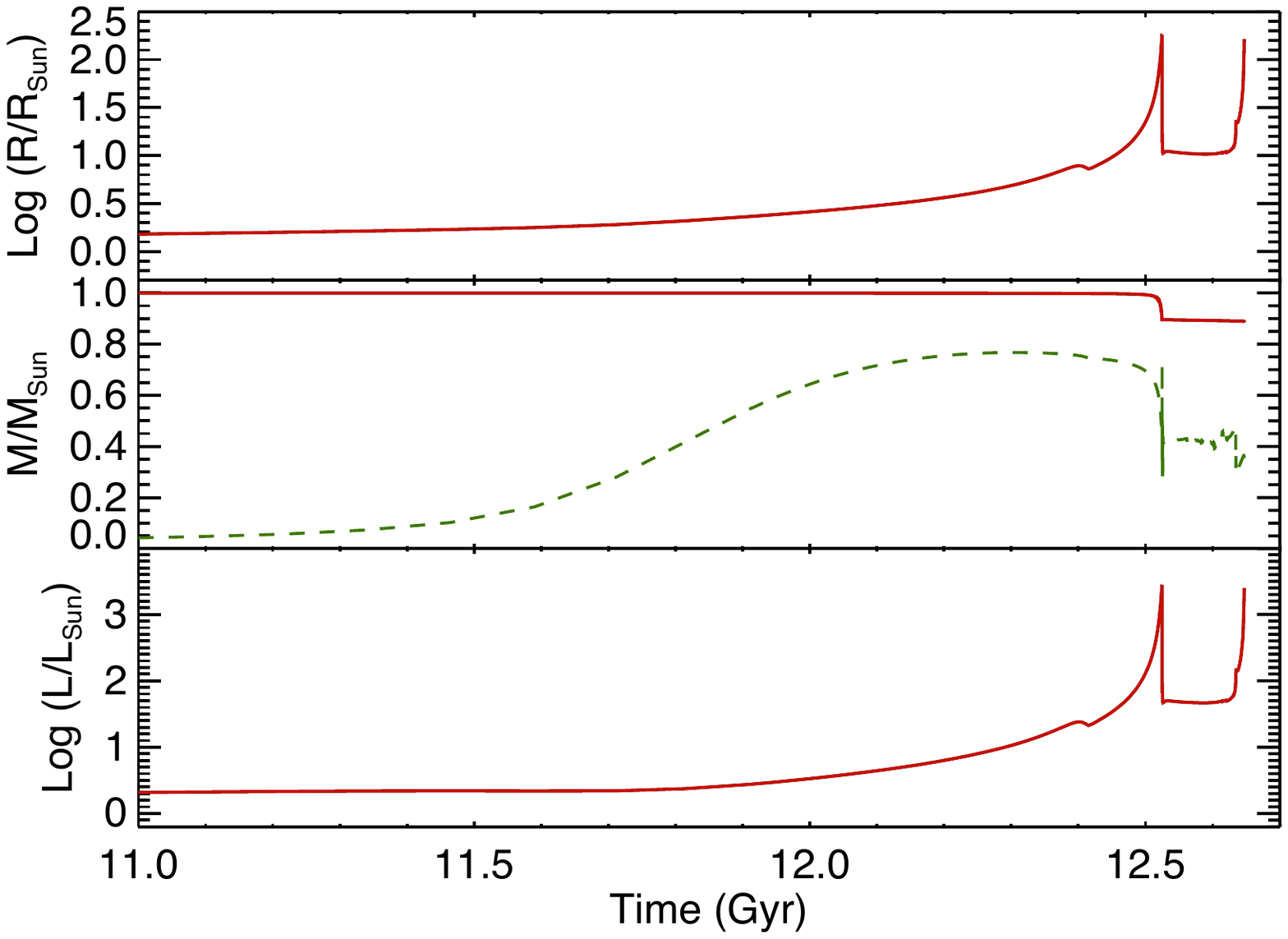} }
\vspace*{-92mm} 
   \caption{Late evolution of the main parameters of our Sun-like stellar model. Top panel: stellar radius (red solid line) vs. the time. Middle panel: stellar mass (red solid line) vs. the time; the mass of the convective zone is overplotted as a green dashed line. Bottom panel: stellar luminosity vs. the time (red solid line). We note the discontinuities in the radius and luminosity when the He flash occurs at the tip of the RGB evolution.  }
              \label{star_evolution}%
\end{figure}

The late evolution of the stellar radius, total mass, mass of the convection zone, and luminosity is shown in Fig.~\ref{star_evolution}. At the tip of the RGB, the maximum radius is $181.8$~R$_{\odot}$, while the  mass and luminosity are 0.896~M$_{\odot}$ and 2776~L$_{\odot}$, respectively. On the other hand, the maximum radius at the tip of the AGB is 163.5~R$_{\odot}$ with a mass of $ 0.886$~M$_{\odot}$ and a luminosity of $\sim 2477$~L$_{\odot}$, respectively. 

\begin{figure}
\vspace*{-9mm}
\hspace*{-15mm}
 \centering{
\includegraphics[width=10cm,height=11.5cm,angle=270]{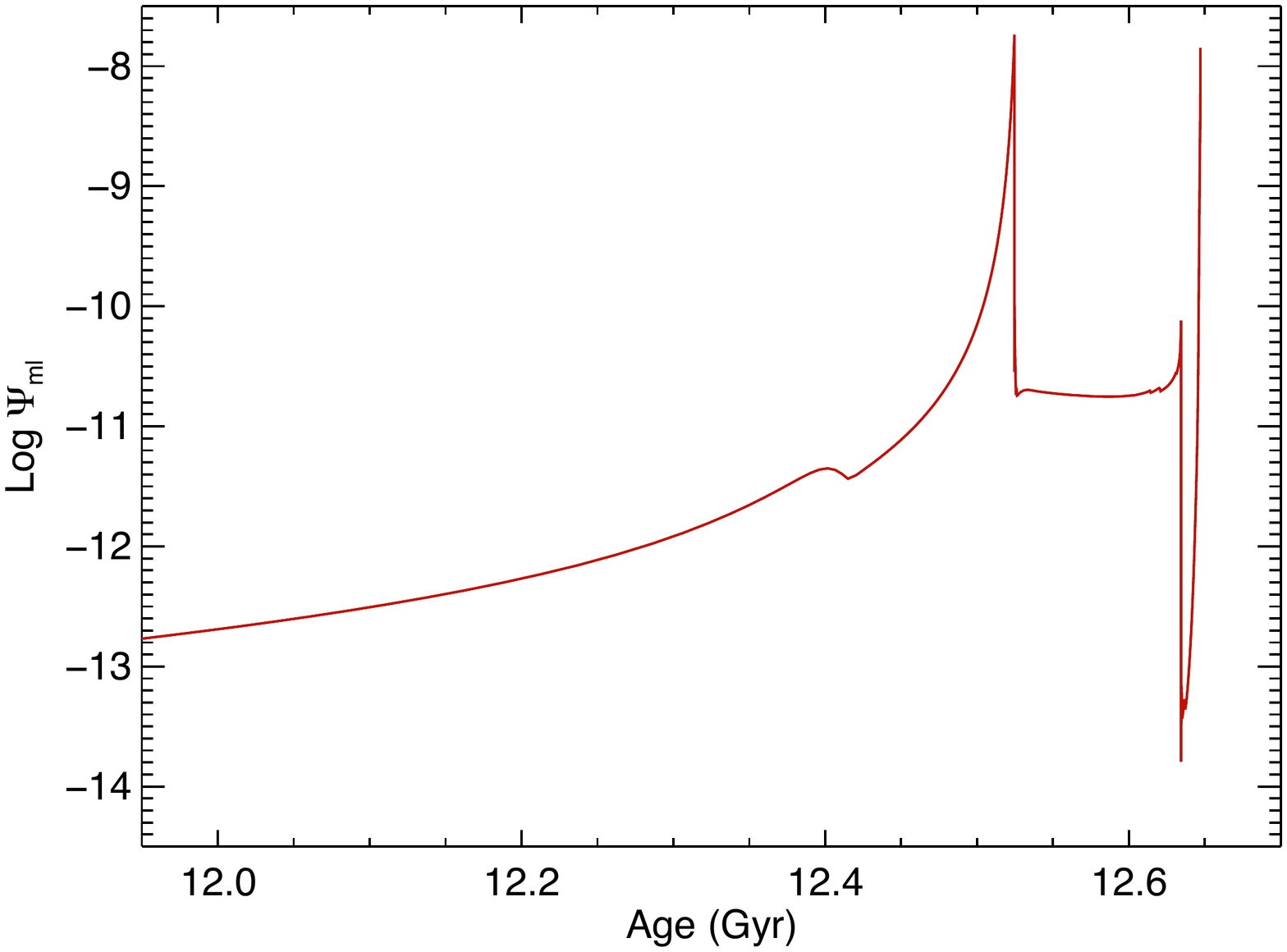}} 
\vspace*{-10mm}
   \caption{The stellar mass-loss parameter $\Psi_{\rm ml}$ vs. the stellar age in the evolutionary phases characterized by a significant stellar mass loss rate. }
              \label{mass_loss_parameter}%
\end{figure}

{  In Fig.~\ref{mass_loss_parameter}, we plot the mass-loss parameter $\Psi_{\rm ml}$ introduced in Eq.~\eqref{Psi_ml_eq} for a planet initially at a distance of 1.5~au from the star, focusing on the late stages of the stellar evolution when $\Psi_{\rm ml}$ reaches its maximal values. We see that $\Psi_{\rm ml}$ is always smaller than $2 \times 10^{-8}$, even at the tip of the RGB or in the AGB phase, when the mass loss rates are the largest. Therefore, we conclude that the adiabatic mass loss approximation is very well verified in our case and our model equations for the evolution of the semimajor axis and the eccentricity given in Sect.~\ref{masslossandtides} hold. We note that in our stellar evolution modeling we always adopt an isotropic stellar mass loss because our stellar structure model is unidimensional.}

We explored several reasonable options for input parameters in MESA, as the use of AGSS09's solar mixture of \citet{2009ARA&A..47..481A}. We found minor differences, of less than $5\%$ in the RGB-tip radius. We also point out that the recent BaSTI-IAC one solar mass model of \citet{2018ApJ...856..125H} including mass loss predicts RGB-tip radii only $3\%$ smaller than ours.

Our stellar parameters differ from those of the evolution model adopted by \citet{SchroederSmith08}, in particular, they find a larger radius of the Sun at the tip of the RGB phase  of $256$~R$_{\odot}$ and a lower mass of $0.668$~M$_{\odot}$. 
We implemented in MESA the \citet{2005ApJ...630L..73S} mass loss formalism used by \citet{SchroederSmith08} both on the RGB and AGB branches. The mass loss is higher and at RGB tip, the mass, luminosity and effective temperature are lower by $21\%$, $1.3\%$ and $95$ K respectively, leading to an RGB tip radius of $\sim191\,R_\odot$, that is $\sim5\%$ higher than our reference value. Furthermore, we point out that in order to better fit the observed position of evolved giants in the Hertzsprung-Russell diagram, \citet{SchroederSmith08} progressively and linearly decreased the value of the mixing-length parameter in their models starting from their solar calibrated value $\alpha_\mathrm{MLT, \odot}=2.0$ at $\log g=1.94$, that is at about half way from the base of the RGB to the RGB tip, down to a value of $1.67$ at $\log g\approx 0.0$ at the RGB tip. We checked that if we mimic this \citet{SchroederSmith08}'s recipe by taking a value of $\alpha_\mathrm{MLT}  \approx \alpha_\mathrm{MLT,\odot} -0.30$ when ascending the RGB, we get higher values of the tip radius and smaller values of the tip $T_\mathrm{eff}$ similar to what these authors got.

\section{Applications}
\label{applications}

We apply our model for the convective-induced residual eccentricity and transit time variation  to the case of a planet of one Earth mass orbiting around a star of initial mass equal to that of the Sun, the structure and evolution of which have been computed as described in Sect.~\ref{stellar_str_evol_model}. 

\begin{figure}
\vspace*{-15mm}
\hspace*{-10mm}
\centering{
\includegraphics[width=11cm,height=15cm,angle=0]{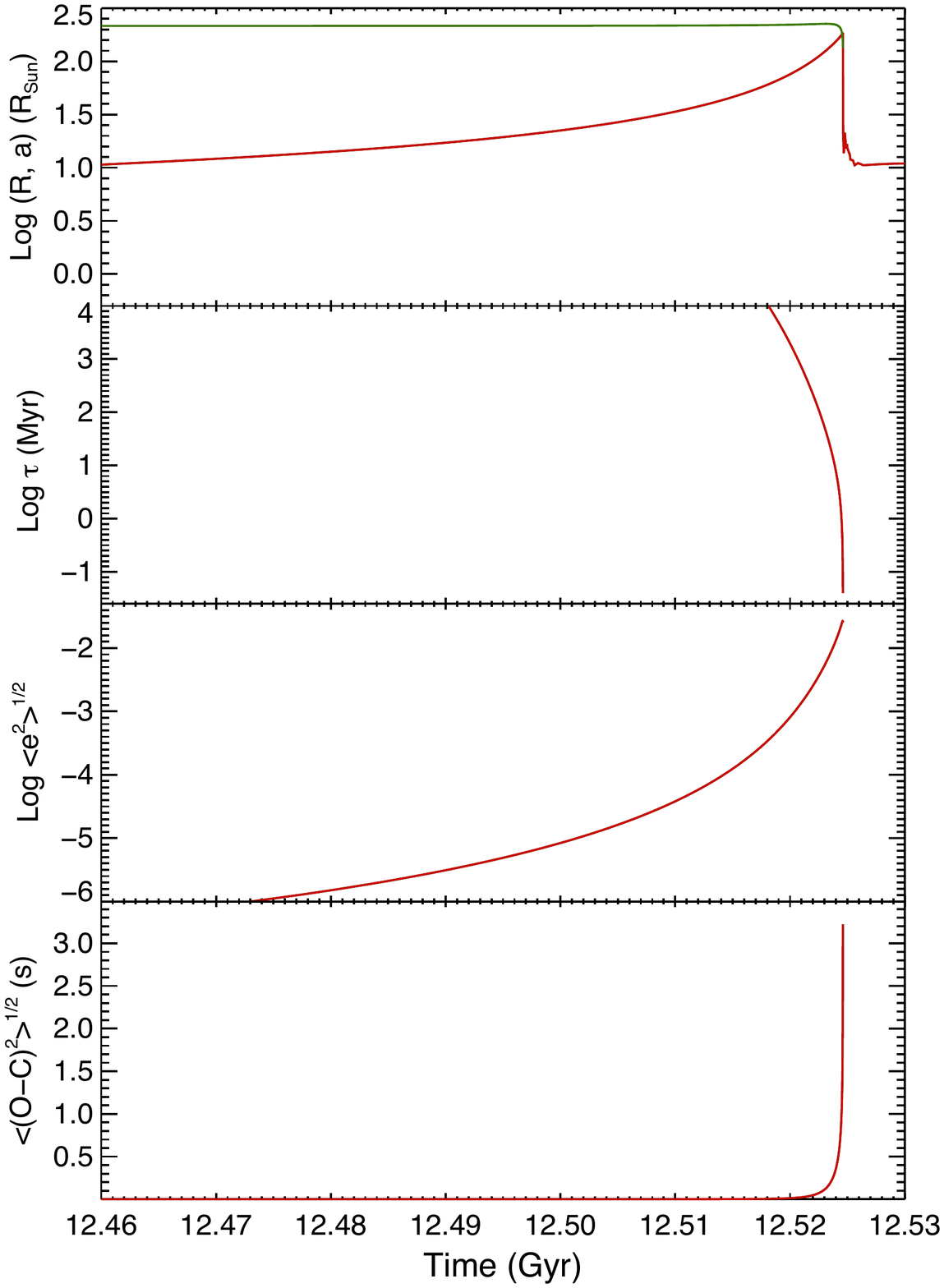}} 
\vspace*{-13mm}
   \caption{Late evolution of the radius of our Sun-like stellar model and orbital parameters of an Earth-mass planet with an initial semimajor axis of 1.0~au. Top panel: stellar radius (red solid line) and orbital semimajor axis (green solid line) vs. the time. Second panel: time scale for the damping of the orbital eccentricity $\tau_{\rm e}$ vs. the time (red solid line) computed according to Eq.~(\ref{tau_ecc}) with $\beta=1$. Third panel: residual eccentricity $\langle e^{2} \rangle^{1/2}$ vs. the time (red solid line) as computed from Eq.~(\ref{residual_e2}). Bottom panel: observed minus calculated time of mid transit $\langle (O-C)^{2} \rangle^{1/2}$ vs. the time (red solid line) computed according to Eqs.~(\ref{o-c2}) and~(\ref{epsilon_o-c2}) over a time interval of ten years. The green line in the top panel and the plots in the second, third, and bottom panels are truncated at the time of the engulfment of the planet by the red giant star. }
              \label{orbit_evolution_1au}%
\end{figure}

In Fig.~\ref{orbit_evolution_1au}, we plot in the top panel the evolution of the stellar radius according to our model and the corresponding change in the orbit semimajor axis for an Earth-mass planet initially at a separation of 1.0~au according to Eqs.~(\ref{tidal_torque_zahn}) and~(\ref{dadt_complete}), where we adopted $\lambda_{2}=0.046$. In our model, the maximum radius at the tip of the RGB phase is 29\% smaller than in \citeauthor{SchroederSmith08}'s 
model, but this is not enough to allow  the Earth to avoid the engulfment, even without including the effect of the residual eccentricity. This happens because the smaller mass loss rate in our model is not capable to overcome the effects of tides leading to the eventual decay of the Earth's orbit. 

{  The fate of the Earth after entering the envelope of the RGB Sun is to slowly spiral toward the center of the star where it will be finally destroyed close to the core at a radius $r_{\rm d}/R \sim 0.003$, where $R$ is the stellar radius at the RGB tip, after a time interval on the order of $\sim 3\times 10^{3}$~yr since its engulfment as we show in Appendix~\ref{final_Earth_fate}. The effects on the stellar structure, luminosity, and rotation are negligible. Even in the case of the engulfment of a giant planet, those effects are not expected to be large given the large luminosity of the giant star and the slow spiral in of the planet \citep[cf.][]{MacLeodetal18,Yarzaetal22}. However, in some systems, the added  effects of successive planetary engulfments or of the additional energy inputs during the helium flash or the AGB thermal pulses may lead to the ejection of the common envelope and the formation of systems such as WD 1856+534 where a white dwarf is accompanied by a planet with a mass of $\la 14$ Jupiter masses and an orbital period of 1.4~days  \citep[e.g.,][]{Lagosetal21,Chamandyetal21,Merlovetal21}}

{  Reverting to the evolution of our planet before its engulfment,} any initial eccentricity of the orbit of the Earth is rapidly erased close to the tip of the RGB because the tidal damping timescale becomes shorter than 0.1~Myr close to that evolution point as shown by the second panel of Fig.~\ref{orbit_evolution_1au}, where $\tau_{\rm e}$ has been computed with Eq.~(\ref{tau_ecc}) adopting $\beta=1$.
Including the effects of the residual eccentricity, the engulfment may occur a little bit earlier because the most probable value of the eccentricity becomes close to 0.01 during the final phase of the RGB ascent. Nevertheless, the difference is of very little practical relevance because the residual eccentricity becomes significant when the orbit of the planet is already on its final tidal decay path (cf. the third panel of Fig.~\ref{orbit_evolution_1au}). The variation in the time of mid transit with respect to a constant-period ephemeris is very small and reaches only a few seconds over a time baseline of ten years during a phase occurring only 1-2~Myr before the engulfment (cf.  the bottom panel of Fig.~\ref{orbit_evolution_1au}). Such a phase is so short in comparison to the evolutionary timescale of the star and the variation in the time of mid transit is so small that it is extremely unlikely to be observable. 

\begin{figure}
\vspace*{-15mm}
\hspace*{-13mm}
\centering{
\includegraphics[width=11.3cm,height=15cm,angle=0]{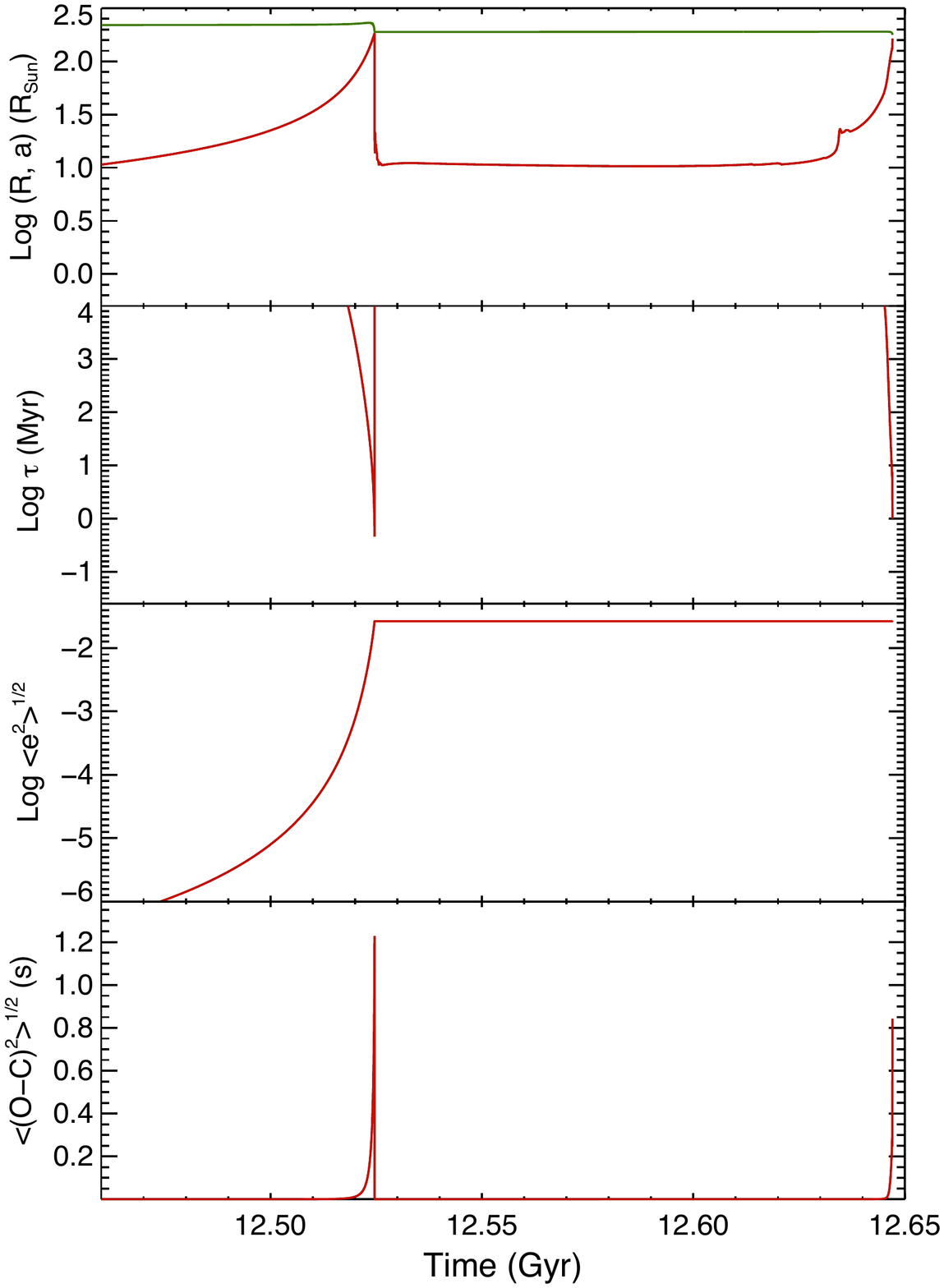}} 
\vspace*{-13mm}
   \caption{Same as Fig.~\ref{orbit_evolution_1au}, but for an Earth-mass planet with an initial semimajor axis of 1.02~au, sufficient to escape engulfment at the tip of the RGB phase when only the tidal orbital decay is considered. Therefore, the plots are extended till the time when the star reaches the AGB tip. }
              \label{orbit_evolution_1.02au}%
\end{figure}

The role of the residual eccentricity becomes relevant if we consider a planet on an initially wider orbit that allows it to escape engulfment. In the top panel of  Fig.~\ref{orbit_evolution_1.02au}, we plot the evolution of the solar radius according to our model and of the orbital semimajor axis for an Earth-mass planet with an initial separation of 1.02~au that is sufficient to avoid engulfment assuming only the tidal orbital decay. This is a consequence of the strong dependence of the tidal decay on the relative orbital separation as expressed by the factor $(R_{\rm s}/a)^{6}$ in Eq.~(\ref{tidal_torque_zahn}). 

Now, when the star reaches the tip of the RGB, the orbit semimajor axis is 1.044 times larger than the radius of the star. The sudden drop in the stellar radius by more than one order of magnitude produced by the internal structure changes associated with the helium flash puts the planet in safety halting its tidal orbital decay. Any initial orbital eccentricity is effectively erased by tides because $\tau_{\rm e}$ becomes shorter than $\sim 1$~Myr close to the RGB tip (cf. the second panel of Fig.~\ref{orbit_evolution_1.02au}).  Nevertheless, the situation changes if we take into account the residual eccentricity because $\langle e^{2} \rangle^{1/2} = 2.65 \times 10^{-2}$ at the tip of the RGB evolution. The planet can avoid engulfment if the eccentricity of its orbit is smaller than 0.044, otherwise its periastron distance becomes smaller than the radius of the star. The probability of such a condition is given by Eq.~(\ref{pe_cumul_prob}) with $e_{0} = 0.044$ and is 0.903. In other words, taking into account the residual eccentricity, there is a $\sim 10$\% probability that the planet be engulfed during its periastron passage close to the tip of the RGB. 

If the planet is capable of escaping engulfment at the RGB tip, its residual eccentricity will be frozen till the end of the stellar evolution because the maximum of the stellar radius reached at the tip of the AGB phase is remarkably smaller than the maximum reached at the RGB tip in our model and does not affect the value of $\langle e^{2} \rangle^{1/2}$ in any significant way. In this case, the eccentricity of the orbit of the planet is maintained over the white-dwarf phase of the system and may play a relevant role in the pollution of the remnant white dwarf by residual planetesimals in the system \citep{FrewenHansen14}. 

The variation in the time of mid transit with respect to a constant-period ephemeris over a time baseline of ten years is plotted in the bottom panel of Fig.~\ref{orbit_evolution_1.02au} and reaches a maximum of 1.2~s. Therefore, it is extremely difficult to detect. 

The maximum residual eccentricity is reached in our model close to the tip of the RGB and remains almost constant throughout the final evolution phases, provided that the planet can avoid engulfment. This is a consequence of the fact that the maximum  stellar radius is reached at the tip of the RGB phase followed by a sudden drop at the helium flash, while the maximum radius reached during the tip of the AGB phase is significantly smaller in our model. Therefore, the damping of the residual eccentricity becomes negligible after the RGB tip -- cf. the dependence of the damping timescale on $(R_{\rm s}/a)$ in Eq.~(\ref{tau_ecc}) -- while the excitation of the eccentricity by the quadrupole moment fluctuations is negligible because they also decrease rapidly with decreasing $(R_{\rm s}/a)$. 

\begin{figure}
\vspace*{-12mm}
\hspace*{-10mm}
\centering{
\includegraphics[width=11cm,height=11.5cm,angle=270]{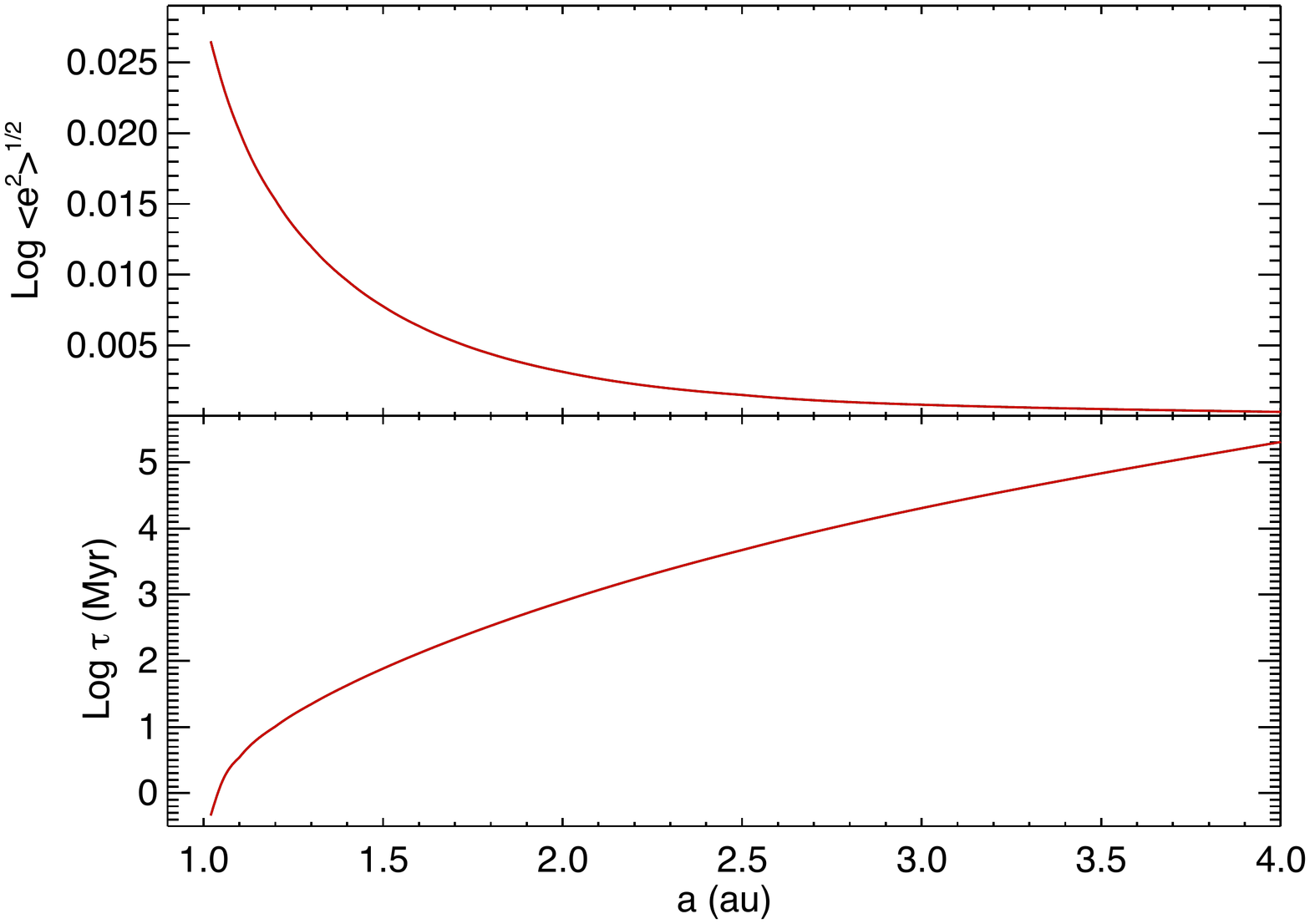}} 
\vspace*{-11mm}
   \caption{Relevant parameters at the tip of the RGB phase vs. the initial semimajor axis of the planetary orbit. Top panel: mean residual eccentricity $\langle e^{2} \rangle^{1/2}$; bottom panel: tidal decay timescale of the eccentricity $\tau_{\rm e}$.}
              \label{res_ecc_vs_a}%
\end{figure}

In Fig.~\ref{res_ecc_vs_a}, we plot the residual eccentricity $\langle e^{2} \rangle^{1/2}$ (top panel) and  the minimum of the eccentricity damping timescale $\tau_{\rm e}$ (lower panel) that are reached at the tip of the RGB phase  for different initial values of the orbital semimajor axis. {  As already noted in Sect.~\ref{perturbations}, $\langle e^{2} \rangle^{1/2}$ is virtually independent of the mass of the planet.} We see that the residual eccentricity at the tip of the RGB and in later evolutionary phases is greater than 0.01 for an initial semimajor axis up to about 1.4~au and decreases rapidly with increasing semimajor axis because of the rapid decrease of the quadrupole potential. Any initial orbital eccentricity is rapidly erased on timescales shorter than $\sim 10$~Myr for an initial semimajor axis $< 1.2$~au leaving only the residual eccentricity in those final phases of the stellar evolution up to the white-dwarf phase. {  On the other hand, for planets with an initial  semimajor axis $a \ga 1.2-1.3$~au, the minimum tidal damping timescale increases remarkably and any initial eccentricity is not erased during the red giant phase of the stellar evolution.  }

{  In planetary systems with a central star in the white dwarf (WD) stage of its evolution, the presence of an eccentric planet is a necessary condition for the perturbation of the orbits of the asteroids leading to their accretion onto the WD. Extensive analytical studies \citep[e.g.,][]{AntoniadouVeras16,AntoniadouVeras19} and numerical simulations \citep[e.g.,][]{FrewenHansen14,Verasetal21} showed that a planet on a circular orbit is not capable of inducing accretion of the asteroids inside the Roche sphere of the WD, while in the case of an eccentric planetary orbit this is possible, even for eccentricities as small as a few hundredths. A detailed exploration of the role played in the WD accretion by the planetary orbit  circularization and the residual eccentricity occurring close to the tip of the RGB is beyond the scope of the present work, but the results summarized in Fig.~\ref{res_ecc_vs_a} can provide a useful starting point for such investigations. } 


\section{Discussion and conclusions}
\label{conclusions}
We introduced a model to compute the residual eccentricity of the orbit of a planet around a star with an outer convective envelope. The eccentricity is maintained against the tidal damping by the random fluctuations of the stellar outer gravitation field due to the small density perturbations that drive convective motions. In the case of an Earth-like planet orbiting a Sun-like star, it is negligible during the main-sequence phase of the evolution of the star, but it becomes relevant during its red giant phase, especially close to the tip of the RGB  when the star becomes hundreds of times larger and thousands of times more luminous than on the main sequence making the amplitude of convective density fluctuations remarkably greater. 

The residual eccentricity is a random variable with a Gaussian  probability density distribution. We find a maximum mean value of the eccentricity $\langle e^{2} \rangle^{1/2} \sim 0.026$ using our model in the case of an Earth-like planet with an initial semimajor axis of 1.02~au, that is, the minimum value to have a probability of $\sim 0.9$ to escape engulfment at the tip of the RGB phase. 

A direct observation of the planet orbit to measure its residual eccentricity requires performing imaging with a contrast on the order of $10^{10}$ at an angular resolution of a microarcsecond to detect an Earth-size planet around a red giant star at a distance of $\sim 1000$~pc. Such observations can become possible, in principle, by using the Sun as a gravitational lens \citep[see][and references therein]{TuryshevToth22}. 

The introduction of the residual eccentricity demonstrates that the engulfment of an Earth-like planet with an initial orbital semimajor axis of 1.0~au by a Sun-like star during its RGB evolution is a stochastic process, in contrast with previous studies  where it was considered to be a deterministic process ruled by the stellar evolution and the tidal decay of the orbit \citep[cf.][]{SchroederSmith08}. 


The late orbital evolution scenario of an Earth-like planet depends critically on the stellar mass loss rate during the red giant and the asymptotic branch phases because it remarkably affects the evolution of the stellar radius and its maximum value. Such a mass loss rate is still poorly known and it may depend on physical processes that are not solely under the control of stellar evolution. Specifically, \citet{SabachSoker18} speculate that the mass loss rate can be increased by a binary companion or a massive close-by planet that produce a remarkable tidal interaction or are engulfed during the ascent of the RGB branch leading to a faster stellar rotation that may enhance the mass loss. Since our mass loss rate parameterizations are calibrated with stars whose binarity status is not known, they suggest that they may suffer a remarkable contamination by binarity effects, while the mass loss rate for a star without massive companions, like our Sun, would be significantly smaller. As a consequence, considering for example a reduction in the mass loss rate by a factor of $\sim 7$, the maximum radius reached at the tip of the AGB branch becomes more than twice larger than that at the tip of the RGB leading to the tidal engulfment of a telluric planet even if its initial orbital semimajor axis is as large as $\sim 1.4-1.5$~au. 

If the planet escapes engulfment during the RGB and AGB phases, the residual eccentricity can play a remarkable role during the white-dwarf phase of a planetary system because it can excite the eccentricity of the orbits of residual planetesimals leading them to collide with the white dwarf and pollute its atmosphere. Even residual eccentricities as small as a few hundredths are sufficient for this \citep[cf.][]{FrewenHansen14}. However, our analysis does not take into account the role of other possible planets in the evolution of the eccentricity of an Earth-like planet surviving the RGB and the AGB phases of its star. 

Angular momentum exchanges among several planets can lead to an increase of the eccentricity of the orbit of an inner low-mass planet as shown, for example, in the angular momentum deficit model by \citet{Laskar97}. We shall not discuss the possible role of other planets in a system because this is outside the scope of the present work, but we note that the residual eccentricity produced by convective fluctuations in a red giant can play a role in the exchanges of angular momentum between planets in a multi-planet system and should be considered in the model of their orbital evolution. 

Another consequence of the fluctuating gravitational potential of a red giant are the small changes in the mean longitude at the epoch of a close-by orbiting planet. Considering a time baseline of ten years, we find deviations  of the time of mid transit with respect to a constant-period ephemeris not exceeding a few seconds. These are completely undetectable because planetary transits are extremely shallow in the case of a red giant star owing to its large radius. Observations in the core of chromospheric spectral lines have been proposed  to improve the observability and the timing of transits across a red giant star ascending the RGB branch because its chromosphere has a radial extension much thinner than the stellar radius, thus enhancing the photometric depth of the transits \citep{Assefetal09}. However, even such a method is not useful in our case because of the extended chromospheres of stars close to the RGB tip and the small radius of Earth-like planets. 

\begin{acknowledgements}
The authors are grateful to an anonymous Referee for a careful reading of the manuscript and several suggestions that helped them to substantially improve the presentation of their work. Research in the field of exoplanetary astronomy at {\it Osservatorio Astrofisico di Catania}, that belongs to  the Italian National Institute for Astrophysics (INAF), is supported by the Italian Ministry for University and Research. 
\end{acknowledgements}

\appendix

\section{Residual eccentricity according to the method of \cite{Phinney92}}
\label{appendix_B}
In this Appendix, we derive the value of $\langle e^{2} \rangle$ by revisiting the approach introduced by \citet{Phinney92} that is based on the equation for the epicyclic motion with respect to an unperturbed circular orbit. We derive the equations of the motion in Sect.~\ref{lagrangian} following the Lagrangian formalism and solve the equation for the radial epicyclic motion in Sect.~\ref{radial_motion} to find the mean squared residual eccentricity. 

\subsection{Lagrangian function of the star-planet system}
\label{lagrangian}

To study the dynamics of our star-planet system, we apply the Lagrangian formalism. The Lagrangian $\cal L$ of our system is defined as
\begin{equation}
{\cal L} = {\cal T} - \Psi,
\end{equation}
where $\cal T$ is the kinetic energy of the system and $\Psi$ its potential energy expressed as functions of the coordinates and their time derivatives in an inertial reference frame. 
We consider a reference frame having its origin at the barycenter $Z$ of the star-planet system and the $x_{0}y_{0}$ plane of which is the orbital plane of the system. Given that $m_{\rm p} \ll m_{\rm s}$, the  barycenter of the system $Z$ virtually coincides with the barycenter $O$ of the star. 
We consider the star as non-rotating and the planet as a point mass $m_{\rm p}$, therefore, we neglect the kinetic energy of their rotation. 

\begin{figure}
 \centering{
\includegraphics[width=10cm,height=6cm,angle=0,trim=87 87 93 93, clip]{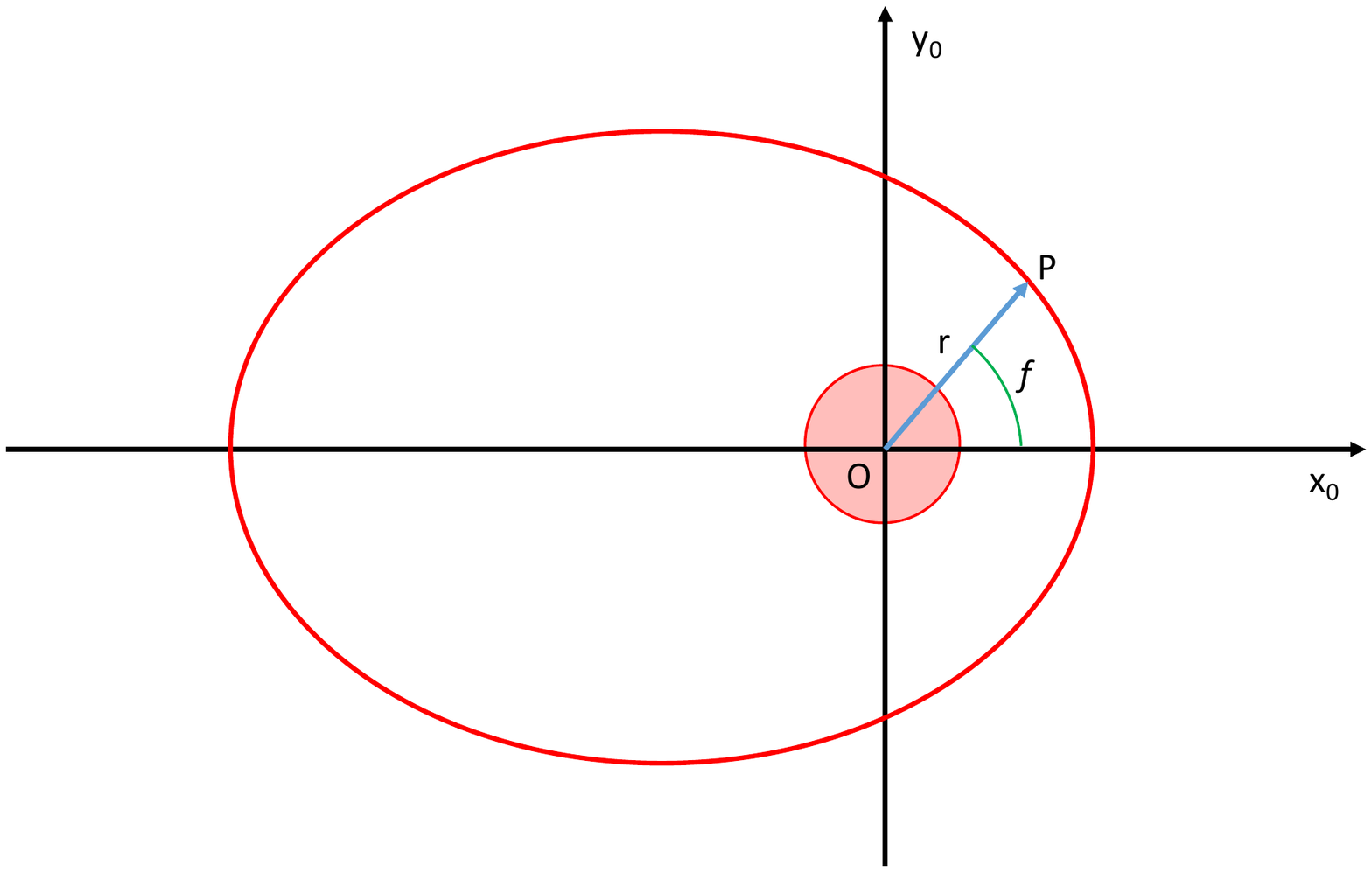}} 
   \caption{Reference frame in the plane of the orbit to describe the relative motion of the planet $P$ around the star and define the true anomaly $f$ and the relative separation of the barycenters of the star and the planet $r = OP$. The origin of the reference frame is at  the barycenter $O$ of the star, while the axis $x_{0}$ points toward the periastron of the orbit. }
              \label{ref_frame}%
\end{figure}

The kinetic energy of the relative orbital motion of our system can be expressed as the energy of the motion of a body having the reduced mass of the system $m = m_{\rm s} m_{\rm p}/(m_{\rm s}+ m_{\rm p})$, where $m_{\rm s}$ is the mass of the star and $m_{\rm p}$ that of the planet, around the barycenter $O$ of the star. Therefore, the expression of the kinetic energy of the system is:
\begin{equation}
{\cal T} = \frac{1}{2} m \left( \dot{r}^{2} + r^{2} \dot{f}^{2} \right), 
\label{kin_ener}
\end{equation}
where $r$ is the relative separation between the two bodies and $f$ is here the true anomaly of the relative orbital motion (see Fig.~\ref{ref_frame}); the dot over a variable indicates its time derivative.  

The gravitational potential energy $\Psi$  is given by (cf. Eq.~\ref{gravit_poten})
\begin{equation}
\Psi  = \Phi  \, m_{\rm p } 
= -\frac{G m_{\rm s} m_{\rm p}}{r} - \frac{3G Q(t) m_{\rm p}}{2 r ^{3}},  
\label{pot_ener}
\end{equation}
where $Q$ is the axisymmetric component of the quadrupole moment of the star as defined in Sect.~\ref{grav_poten}; because of the density fluctuations, it is a function of the time $t$. 


The equations of motion of our system can be derived from its Lagrangian as
\begin{equation}
\frac{d}{dt} \left( \frac{\partial {\cal L}}{\partial \dot{q}}  \right)- \frac{\partial {\cal L}}{\partial q} = 0,
\end{equation}
where $t$ is the time and $q = r, f $ is any of the coordinates adopted to describe the motion of the system. In this way, we obtain the following equations of motion:
\begin{eqnarray}
 m \ddot{r} -mr \dot{f}^{2}   +  \frac{Gm_{\rm s} m_{\rm p}}{r^{2}}  +  \frac{9G Q(t) \, m_{\rm p}}{2 r^{4}} & = & 0, \label {r_eq} \\ 
m \frac{d}{dt} \left( r^{2} \dot{f} \right) & = & 0. \label{f_eq} 
\end{eqnarray}
Equation~(\ref{f_eq}) can be immediately integrated to give the conservation of the orbital angular momentum $J$ of the system 
\begin{equation}
m r^{2} \dot{f} = J. 
\label{ang_mom_conserv}
\end{equation}
Considering the definition of the reduced mass, the equation of the radial motion becomes:
\begin{equation}
\ddot{r} - r \dot{f}^{2} + \frac{Gm_{\rm T}}{r^{2}} + \frac{9G m_{\rm T} Q(t)}{2 m_{\rm s} r^{4}} = 0, 
\label{rnew_eq}
\end{equation}
where $m_{\rm T} \equiv m_{\rm s} + m_{\rm p}$ is the total mass of the system.  

\subsection{Radial motion}
\label{radial_motion}

We consider an orbit of small eccentricity, so that
\begin{equation}
r(t) = a [1+ x(t)],
\label{x_def}
\end{equation}
where $x(t)$  is the relative deviation from a circular orbit with $ | x(t) | \ll 1$. 
Making use of the conservation of the total angular momentum $J$ as given by Eq.~(\ref{ang_mom_conserv}), the equation of the radial motion (\ref{rnew_eq}) becomes:
\begin{equation}
\ddot{r} - \frac{J^{2}}{m^{2} r^{3}} + \frac{Gm_{\rm T}}{r^{2}} + \frac{9GQ(t) m_{\rm T}}{2 m_{\rm s} r^{4}}  = 0.  
\label{rad_eq_compl}
\end{equation}
Substituting Eq.~(\ref{x_def}) into Eq.~(\ref{rad_eq_compl}) and considering that $\ddot{r} = a \ddot{x}$ and $(1+x)^{-q} \simeq (1-q x)$, we find: 
\begin{equation}
 a \ddot{x} - \frac{J^{2}}{m^{2} a^{3}} (1 -3x) + \frac{Gm_{\rm T}}{a^{2}} (1-2x) + \frac{9 G Q m_{\rm T}}{2 m_{\rm s} a^{4}} (1-4x) = 0.
\end{equation}
Making use of fact that $J \simeq m n a^{2}$ for $e\ll 1$ (cf. Eq.~\ref{orbital_ang_mom}), the Kepler third law, and collecting all the terms in $x$, we obtain to the first order:
\begin{equation}
\ddot{x} + n^{2}\left(1 - \frac{18 Q}{m_{\rm s} a^{2}} \right) x = - \frac{9}{2} n^{2}\frac{Q}{m_{\rm s} a^{2}}. 
\end{equation}
The second term in brackets on the left-hand side is very small in comparison to the first because the quadrupole moment fluctuations $| Q |\ll m_{\rm s} a^{2}$. In conclusion, we can write the equation of the epicyclic motion as:
\begin{equation}
\ddot{x} + n^{2}  x \simeq - \frac{9}{2} n^{2} \frac{Q (t)}{m_{\rm s} a^{2}}, 
\label{epicyc_eq0}
\end{equation}
that is the equation of a forced harmonic oscillator of frequency $n$ equal to that of the orbital motion. The time-dependent nature of the forcing has been made explicit by indicating that $Q$ is a function of the time $t$. The relationship between $x$ and the eccentricity $e$ comes from the solution of Eq.~\eqref{epicyc_eq0}, that is, $x= e \sin (nt)$. 

Tides inside the star and the planet damp any initial orbital eccentricity over a characteristic timescale $\tau_{\rm e}= e/ |de/dt |$ that can be calculated from the tidal theory (cf. Eq.~\ref{tau_ecc}). Therefore, Eq.~(\ref{epicyc_eq0}) must be completed by adding a term that takes into account the damping of the radial motion by the tides because they tend to restore a circular orbit. Such a term can be expressed as a dissipative term in the equation for the radial oscillations that becomes:
\begin{equation}
\ddot{x} + 2b \dot{x}+ n^{2}  x \simeq - \frac{9}{2} n^{2} \frac{Q (t)}{m_{\rm s} a^{2}}. 
\label{epicyc_eq}
\end{equation}
 This is the equation of a damped harmonic oscillator with a random forcing because $Q (t)$ is a stochastic function of the time. Its solution, for $b \ll n$, is $x \sim \exp(-bt) \sin (nt)$, that can be compared with the undamped solution $x= e \sin(nt)$ to give the relationship between $b$ and $\tau_{\rm e}$, that is, $b = \tau_{\rm e}^{-1}$ \citep[cf.][\S~25]{LandauLifshitz69}. 

Taking the Fourier transform of both sides of Eq.~(\ref{epicyc_eq}) with $i=\sqrt{-1}$, we obtain
\begin{equation}
- \omega^{2} \tilde{x}(\omega) + 2b i \, \omega \tilde{x}(\omega) + n^{2} \tilde{x} (\omega) = - \frac{9}{2} n^{2} \frac{\tilde{Q} (\omega)}{m_{\rm s} a^{2}},
\end{equation}
where the tilde indicates the Fourier transform and $\omega$ is the frequency. This can be recast as
\begin{equation}
\tilde{x}(\omega) =  - \frac{9}{2} n^{2} \frac{\tilde{Q} (\omega)}{m_{\rm s} a^{2}} \frac{1}{Z(\omega)}, 
\end{equation}
where the function 
\begin{equation}
Z(\omega) = \frac{1}{ n^{2}-\omega^{2} + 2b i  \, \omega}. 
\end{equation}
The power spectrum $P_{x}(\omega) \equiv \tilde{x}(\omega) \tilde{x}^{*}(\omega) $ of the solution $x(t)$ can be immediately computed as
\begin{multline}
P_{x}(\omega) = \frac{81}{4} n^{4} \frac{P_{Q}(\omega)}{m_{\rm s}^{2} a^{4}} \frac{1}{Z(\omega) Z^{*}(\omega)} = \\
= \frac{81}{4} n^{4} \frac{P_{Q}(\omega)}{m_{\rm s}^{2} a^{4}} \frac{1}{(n^{2} - \omega^{2})^{2} + 4b^{2}\omega^{2}},
\end{multline}
where the asterisk indicates complex conjugation and $P_{Q} \equiv \tilde{Q}(\omega) \tilde{Q}^{*}(\omega)$ is the power spectrum of the quadrupole fluctuations. The expectation value of $x^{2}$, that is, $\langle x^{2} \rangle$, given that $\langle x \rangle = 0$, is given by
\begin{multline}
\langle x^{2} \rangle = \frac{1}{2\pi} \int_{-\infty}^{\infty} P_{x} (\omega)\, d\omega = \\
= \frac{81}{8\pi} n^{4} \frac{1}{m_{\rm s}^{2} a^{4}} \int_{-\infty}^{\infty} \frac{P_{Q}(\omega)}{(n^{2} - \omega^{2})^{2} + 4b^{2} \omega^{2}} \, d\omega. 
\end{multline}
Because $b \ll n$, the integrand is a very sharply peaked function around  $\omega \sim n$, that is, only the power spectrum of the quadrupole fluctuations very close to the resonance contributes to the integral. Therefore, without any significant loss of accuracy, we can write
\begin{equation}
\langle x^{2} \rangle 
\simeq \frac{81}{8\pi} n^{4} \frac{1}{m_{\rm s}^{2} a^{4}}\, P_{Q} (n) \int_{-\infty}^{\infty} \frac{d \omega}{(n^{2} - \omega^{2})^{2} + 4b^{2}\omega^{2}}.   
\label{ave_x2}
\end{equation}
The integral can be computed, for example, by means of the method of the residues of the theory of complex variable and its value is \citep[cf. Appendix of][]{Lanza21}:
\begin{equation}
\int_{-\infty}^{\infty} \frac{d \omega}{(n^{2} - \omega^{2})^{2} + 4b^{2}\omega^{2}} = \frac{\pi}{2bn^{2}}.
\label{notable_int}
\end{equation}
Making use of Eq.~(\ref{notable_int}), expression (\ref{ave_x2}) becomes
\begin{equation}
\langle x^{2} \rangle = \frac{81}{16}  \frac{n^{2}}{b} \frac{P_{Q}(n)}{m_{\rm s}^{2} a^{4}}. 
\end{equation}
The expectation value of the squared velocity $\langle \dot{x}^{2} \rangle$ can be similarly obtained by taking into account that $\tilde{\dot{x}} = i\omega \tilde{x} (\omega) $ so that 
\begin{equation}
P_{\dot{x}} (\omega) = \omega^{2} P_{x} (\omega). 
\end{equation}
The analogous equation to Eq.~(\ref{ave_x2}) then is
\begin{multline}
\langle \dot{x}^{2} \rangle = \frac{1}{2\pi} \int_{-\infty}^{\infty} \omega^{2} P_{x} (\omega) \, d\omega \simeq  \\
\simeq \frac{81}{8\pi} n^{4} \frac{1}{m_{\rm s}^{2} a^{4}} P_{Q} (n) \int_{-\infty}^{\infty} \frac{\omega^{2} d \omega}{(n^{2} - \omega^{2})^{2} + 4b^{2}\omega^{2}}.
\end{multline}
Applying again the method of the residues, the integral is found to be \citep[cf. Appendix of][]{Lanza21}
\begin{equation}
\int_{-\infty}^{\infty} \frac{\omega^{2} d \omega}{(n^{2} - \omega^{2})^{2} + 4b^{2}\omega^{2}} = \frac{\pi}{2b}.
\end{equation}
In conclusion
\begin{equation}
\langle \dot{x}^{2} \rangle = \frac{81}{16} \frac{n^{4}}{b} \frac{P_{Q} (n)}{m_{\rm s}^{2} a^{4}}. 
\label{ave_xdot2}
\end{equation}
Therefore, the average mechanical energy of the epicyclic oscillator is
\begin{equation}
\langle E \rangle \equiv \frac{1}{2} m a^{2} \langle \dot{x}^{2} \rangle + \frac{1}{2} m n^{2} a^{2} \langle x^{2} \rangle = \frac{81}{16}\, m \, \frac{n^{4}}{b} \frac{P_{Q} (n)}{m_{\rm s}^{2} a^{2}}, 
\label{eave_pow}
\end{equation}
where $m \simeq m_{\rm p}$ is the reduced mass of the system. 

As we saw before, the stationary solution of the equation for the epicyclic radial motion can be approximated as $x(t) = e(t) \sin (n t)$, where  $e \ll 1$ is the eccentricity of the orbit. Because $b\ll n$, the timescale upon which $e$ varies is remarkably longer than the orbital period. Therefore, $\langle x^{2} \rangle = (1/2) \langle e^{2} \rangle $, $\langle \dot{x}^{2} \rangle = (1/2) n^{2} \langle e^{2} \rangle $, and the average value of the total mechanical energy is: 
\begin{equation}
\langle E \rangle = \frac{1}{4} \, m \, n^{2} a^{2} \langle e^{2} \rangle + \frac{1}{4}\, m \, n^{2} a^{2} \langle e^{2}  \rangle = \frac{1}{2} \, m \, n^{2} a^{2} \langle e^{2} \rangle.
\label{eave_osc}
 \end{equation}
 Comparing Eq.~(\ref{eave_osc}) with Eq.~(\ref{eave_pow}), we find the average value of the squared residual eccentricity induced by the convective fluctuations of  the quadrupole moment of the star 
  \begin{equation}
 \langle e^{2}  \rangle = \frac{81}{8} \frac{n^{2}}{b} \frac{P_{Q} (n)}{m_{\rm s}^{2} a^{4}} =  \frac{81}{8}  \frac{n^{2} \tau_{\rm e }P_{Q} (n)}{m_{\rm s}^{2} a^{4}}.
 \label{res_e}
 \end{equation}
 It is interesting to compare this result with that given by Eq.~\eqref{residual_e2} for a constant $\tau_{\rm e}$. Developing the exponential into a Taylor series and considering only the leading order term, the expression derived from Eq.~\eqref{residual_e2} is the same as Eq.~\eqref{res_e} with $\tau_{\rm e}$ replaced by $t$, that is:
  \begin{equation}
 \langle e^{2}  \rangle \simeq  \frac{81}{8}  \frac{n^{2} P_{Q} (n)\, t}{m_{\rm s}^{2} a^{4}}.
 \label{res_e_1st}
 \end{equation}
 Therefore, Eq.~\eqref{res_e} is the leading order term of the exact solution as given by Eq.~\eqref{residual_e2} provided that we assume $t=\tau_{\rm e}$.
 Among the hypotheses implicit in the simplified approach adopted to derive Eq.~\eqref{res_e} is the constancy of $\tau_{\rm e}$ and the consideration of a weak eccentricity damping that is valid for $t \ll \tau_{\rm e}$ which is the same for the validity of the truncation of the general solution \eqref{residual_e2} to the leading order. 
 
 Equation~\eqref{res_e_1st} shows that $\langle e^{2} \rangle$ increases proportionally to the time as expected in the case of a Brownian motion. However, the maximum timescale for which such a model is valid is $t \la \tau_{\rm e}$ because tides erase the eccentricity on timescales comparable with or longer than $\tau_{\rm e}$ losing memory of its previous accumulation steps. Therefore, regarding $\tau_{\rm e}$ as the memory timescale for the growth of $\langle e^{2} \rangle$, Eq.~\eqref{res_e_1st} reproduces the result of Eq.~\eqref{res_e}.  
 
 The power dissipated by the damping of the radial oscillations due to the tides can be derived by multiplying Eq.~(\ref{epicyc_eq}) by $m a^{2} \dot{x}$ and writing the resulting equation as
 \begin{equation}
 \frac{dE}{dt} = -2\, b\, ma^{2} \dot{x}^{2} - \frac{9}{2} n^{2} m a^{2} \dot{x} \, \frac{Q(t)}{m_{\rm s} a^{2}}. 
 \end{equation}
 The dissipated power is given by the first term on the right-hand side of this equation because the other term represents the power of the forcing that maintains the oscillations. This implies that the average dissipated power per oscillation cycle $P_{\rm d}$ is
 \begin{equation}
 P_{\rm d} \equiv - \langle \frac{dE}{dt} \rangle = 2b\, ma^{2} \langle \dot{x}^{2} \rangle,
 \end{equation}
 that becomes
 \begin{equation}
 P_{\rm d} = \frac{81}{8} \frac{m \, n^{4}}{m_{\rm s}^{2} a^{2}} P_{Q} (n),
 \label{pdiss}
 \end{equation}
 thanks to Eq.~(\ref{ave_xdot2}); the value of $P_{Q}(n)$ is given by Eq.~\eqref{power_q(n)}. In our model, tidal dissipation occurs inside the star, therefore, such a power makes a very small addition to the stellar luminosity. In the stationary regime, it is independent of the eccentricity damping timescale and scales as $a^{-8}$ as can be seen by substituting from the Kepler third law into Eq.~(\ref{pdiss}). 
 
\section{Statistical distribution of the residual eccentricity}
\label{appendix_A}

The statistical distribution of the residual eccentricity can be computed following the same method used to derive the statistical distribution of the velocities of the molecules in an ideal gas, the so-called Maxwell-Boltzmann distribution. 

Since the direction of the periastron in the orbital plane is randomly oriented, the distribution of the argument of periastron $\varpi_{\rm p}$ of the planetary orbit is uniform in $[0, 2\pi]$. In other words, the distributions of the stochastic variables
\begin{equation}
\left\{
\begin{array}{ccc}
e_{x} & \equiv & e \cos \varpi_{\rm p}, \\
e_{y} & \equiv & e \sin \varpi_{\rm p},  
\end{array} 
\right.
\end{equation}
are the same because there is no preferred orientation of the line of the apsides. We indicate such a distribution as $\varphi(e_{x}) = \varphi(e_{y})$. This distribution must be symmetric, that is, $\varphi(z) = \varphi(-z)$, where $z = e_{x}$ or $e_{y}$; therefore, it will depend on $e_{x}^{2}$ or $e_{y}^{2}$. 

The probability that the eccentricity vector ${  e} \equiv (e_{x}, e_{y})$ has components between $e_{x}$ and $e_{x} + de_{x}$ and $e_{y}$ and $e_{y}+de_{y}$
is $\varphi(e_{x}^{2}) \, \varphi(e_{y}^{2})\, de_{x} \, de_{y}$. Such a probability must depend only on the square of the eccentricity $e^{2} = e_{x}^{2} + e_{y}^{2}$, hence it can be expressed as
$f_{\rm e}(e^{2}) \, de_{x} \, de_{y}$, where $f_{\rm e}$ is the distribution function of $e$ that depends on $e^{2}$. In conclusion, we find
\begin{equation}
f_{\rm e} (e_{x}^{2} + e_{y}^{2}) = \varphi(e_{x}^{2}) \, \varphi(e_{y}^{2}).  
\label{f-ecc}
\end{equation} 
By taking the logarithms and deriving both sides of Eq.~(\ref{f-ecc}) with respect to $e_{x}^{2}$ and $e_{y}^{2}$, respectively, we find
\begin{eqnarray}
\frac{f_{\rm e}^{\prime}(e^{2})}{f_{\rm e}(e^{2})} & = & \frac{\varphi^{\prime}(e_{x}^{2})}{\varphi(e_{x}^{2})}, \label{log_der1} \\
\frac{f_{\rm e}^{\prime}(e^{2})}{f_{\rm e}(e^{2})} & = & \frac{\varphi^{\prime}(e_{y}^{2})}{\varphi(e_{y}^{2})}.
\label{log_der2}
\end{eqnarray}
Equation~(\ref{log_der1}) tells us that $f_{\rm e}^{\prime}/f_{\rm e}$ is independent of $e_{y}^{2}$, while Eq.~(\ref{log_der2}) tells us that it is independent of $e_{x}^{2}$. Therefore, $f_{\rm e}^{\prime}/f_{\rm e} $ must be equal to a constant,  that we denote with $-\zeta$, where $\zeta > 0$ for normalization reasons
\begin{equation}
f_{\rm e}^{\prime} (e^{2}) = -\zeta f_{\rm e}(e^{2}). 
\end{equation}
This equation can be immediately integrated to give
\begin{equation}
f_{\rm e}(e^{2}) \propto \exp(-\zeta e^{2}). 
\end{equation}
The distribution function must be normalized as  
\begin{equation}
\int_{0}^{1} f_{\rm e}(e^{2}) \, de = 1. 
\label{f-norm}
\end{equation}
On the other hand, the mean value of the square of the eccentricity is
\begin{equation}
\langle e^{2} \rangle = \int_{0}^{1} e^{2} f_{\rm e}(e^{2}) \, de.
\label{e2_from_fe}
\end{equation}
Equations (\ref{f-norm}) and (\ref{e2_from_fe}) can be used to find the normalization constant $A = \left( \int_{0}^{1} \exp(-\zeta e^{2})\, de \right)^{-1}$ and express $\zeta$ in terms of $\langle e^{2} \rangle$. This can be done considering that $\langle e^{2} \rangle \sim 10^{-5} - 10^{-4}$ is always very small, so the exponential becomes virtually zero well before reaching the upper limit of integration $e=1$. In other words, we can extend the upper limit of integration to $+\infty$ without any appreciable error and make use of the notable integrals
\begin{equation}
\int_{0}^{\infty} \exp(-\zeta x^{2}) \, dx = \frac{1}{2} \sqrt{\frac{\pi}{\zeta}}, 
\end{equation}
and
\begin{equation}
\int_{0}^{\infty} x^{2} \exp(-\zeta x^{2}) \, dx = \frac{1}{4} \sqrt{\frac{\pi}{\zeta^{3}}}
\end{equation}
to find 
\begin{eqnarray}
A & = & \sqrt{\frac{2}{\pi \langle e^{2} \rangle}}, \\
\zeta & = & \frac{1}{2 \langle e^{2} \rangle}, 
\end{eqnarray}
that give the distribution function of the residual eccentricity as
\begin{equation}
f_{\rm e}(e^{2}) = \sqrt{\frac{2}{\pi \langle e^{2} \rangle}} \exp \left( - \frac{e^{2}}{2\langle e^{2} \rangle}\right). 
\end{equation}
In other words, it is a Gaussian distribution with zero mean and standard deviation $\langle e^{2} \rangle^{1/2}$.  
 
\section{On the final fate of the Earth inside the RGB solar envelope}
\label{final_Earth_fate}

{  When the Earth enters the envelope of the red giant Sun, it will experience an hydrodynamical drag and a gravitational drag. The ratio of the former to the latter is on the order of $(\varv_{\rm orb}/\varv_{\rm esc})^{2} \sim 10$, where $\varv_{\rm orb}$ is the orbital velocity and $\varv_{\rm esc}$ the escape velocity at the surface of our planet \citep[cf. Sect.~2.1 of][]{Yarzaetal22}. Therefore, we neglect the gravitational drag and consider only the hydrodynamical drag given by $F_{\rm D} = C_{\rm D} \pi \, \rho \, \varv_{\rm orb}^{2} \, R_{\rm e}^{2}$, where $C_{\rm D}$ is the drag coefficient of order unity, $\rho$ the density in the stellar envelope at the orbital radius of the planet, and $R_{\rm e}$ the radius of the Earth. 

According to \citet{JiaSpruit18}, the disruption of the Earth takes  place at the disruption radius $r_{\rm d}$ where the drag pressure becomes equal to the density of the  gravitational binding energy of the Earth, that is, where the ratio 
\begin{equation}
f \equiv \frac{\rho \varv_{\rm orb}^{2}}{\rho_{\rm e} \varv_{\rm esc}^{2}} \sim 1,
\end{equation}
with $\rho_{\rm e} = 3m_{\rm e}/(4\pi R_{\rm p}^{3})$ being the mean density of the Earth and $m_{\rm e}$ the Earth's mass. Once such a condition is reached, the planet starts to fragment into smaller and smaller pieces over a characteristic fragmentation timescale $t_{\rm d} \sim P_{\rm orb} \sqrt{\rho_{\rm d}/\rho_{\rm e}}$, where $\rho_{\rm d}$ is the mean stellar density inside the disruption radius and $P_{\rm orb}$ the orbital period around the center of the Sun.  

Considering our internal structure model of the Sun at the tip of the RGB phase, we estimate that the disruption radius is $r_{\rm d}/R \sim 0.003$, where $R$ is the radius of the star, because of the very small density of the stellar plasma in the extended convective envelope. The disruption timescale is very short ($t_{\rm d} \sim 800$~s), therefore, when the condition $f \sim 1$ is reached, the Earth will be rapidly fragmented and dissolved in the Sun's envelope \citep[cf. Sect.~5.1 of][]{JiaSpruit18}. We note that thermal evaporation and ablation play a secondary role in the destruction of the Earth because the density in the solar red giant envelope is much smaller than the mean density of the Earth \citep[see Sect.~2.4 of][for details]{JiaSpruit18}. 

The decay of the Earth's orbit is ruled by the variation in its orbital angular momentum $J$ that can be written as
\begin{equation}
J = m a^{2} n,
\end{equation}
where $m \sim m_{\rm e}$ is the reduced mass of the Earth orbiting the Sun, $a$ the orbit semimajor axis, and $n= 2\pi/P_{\rm orb}$ the mean orbital motion. Making use of the Kepler third law, we derive a differential equation for the evolution of the semimajor axis under the effect of the drag, assumed to be tangential to the circular orbit
\begin{equation}
\frac{da}{dt} = - 2\pi \, C_{\rm D} \, \frac{G(M+m_{\rm e})^{1/2}}{m} \, R_{\rm e}^{2} \, \rho \, a^{1/2}. 
\label{dadt_appC}
\end{equation}
In the layers where the Earth orbits before its disruption, that is, where $r \geq r_{\rm d}$, we approximate the density with a power law as $\rho (r) = \rho(R) (r/R)^{q}$, where $\rho(R)$ is the density at the surface of the star where $r=R$ and $q$ is a fixed exponent. Such an approximation is valid within one order of magnitude, therefore, it is suitable for our approximate treatment of the problem. By fitting the internal density stratification in the model of the Sun at the RGB tip for $r \ge r_{\rm d}$, we find $q=-2.13$. 

Making use of the above approximation for the density and considering that $r_{\rm d} \ll R$, we integrate analytically Eq.~\eqref{dadt_appC} and find the timescale $t_{\rm p}$ of the Earth spiral-in from the engulfment at the surface $r=R$ down to the disruption radius $r=r_{\rm d}$ as
\begin{equation}
t_{\rm p} \sim \frac{2}{1-2q} \left\{ 2\pi C_{\rm D} \left[\frac{G(M+m_{\rm e})}{R}\right]^{1/2} \frac{1}{m} \, \rho(R) \, R_{\rm e}^{2}\right\}^{-1}.
\end{equation}
Adopting $C_{\rm D} = 1$, we find $t_{\rm p} \sim 3 \times 10^{3}$~years because the Earth's spiral-in is initially very slow owing to the very low density in the extended solar envelope. 

The dissipation rate of the orbital energy $E_{\rm orb}$ due to the drag force can be estimated by assuming that the mass inside the orbit of the planet is constant and equal to the total mass of the star as 
\begin{equation}
\frac{dE_{\rm orb}}{dt} = G \, \frac{Mm_{\rm e}}{2 a^{2}} \, \left(\frac{da}{dt} \right),
\end{equation}
where $da/dt$ is given by Eq.~\eqref{dadt_appC}. 

The minimum dissipation rate is that at the engulfment when $a=R$, and it is only $\sim 10^{-8}$ of the stellar luminosity at the tip of the RGB phase. Immediately before the disruption, $a=r_{\rm d}$ and  the maximum of the dissipation rate is reached that amounts to $\sim 10$ times the stellar luminosity. However, such an additional energy input is limited to a short burst occurring on a timescale on the order of a few times the shortest orbital period, that is, a few days at most. The excess heat is redistributed by convection over the entire convective envelope on a timescale $R^{2}/\eta_{\rm turb} \sim 200$~yr, where $\eta_{\rm turb}$ is the turbulent diffusivity of the envelope that is estimated as the average of the product of the mixing length by the convective velocity over the envelope itself. Such an excess heat is finally radiated at the stellar surface on a timescale comparable with the Kelvin-Helmoltz timescale of the envelope itself, that is,  on the order of $\sim 25$~years. Therefore, its impact on the stellar luminosity is negligible. 

}

\end{document}